\documentclass[11pt]{article}
\usepackage[top=1in,bottom=1in,left=1in,right=1in]{geometry}
\usepackage{authblk}
\usepackage{blindtext}
\usepackage{amsmath,amssymb,amsthm}
\usepackage[T1]{fontenc}
\usepackage[tt=false]{libertine}
\usepackage[scaled=0.83]{beramono}
\usepackage[libertine]{newtxmath}
\usepackage[usenames,dvipsnames]{xcolor}
\usepackage{graphicx}
\usepackage{bm}
\usepackage{array,multirow}
\usepackage{booktabs}
\usepackage{mathtools}
\mathtoolsset{showonlyrefs}
\usepackage{definitions_with_apxproof}

\allowdisplaybreaks
\begin{document}

\noeqref{n_1_direct_sum_1-apx}
\noeqref{n_2_direct_sum_1-apx}
\noeqref{n_3_direct_sum_1-apx}
\noeqref{n_1_2_direct_sum_1-apx}
\noeqref{n_1_3_direct_sum_1-apx}
\noeqref{n_2_3_direct_sum_1-apx}
\noeqref{n_1_direct_sum_1_2-apx}
\noeqref{n_2_direct_sum_1_2-apx}
\noeqref{n_3_direct_sum_1_2-apx}
\noeqref{n_1_2_direct_sum_1_2-apx}
\noeqref{n_1_3_direct_sum_1_2-apx}
\noeqref{n_2_3_direct_sum_1_2-apx}
\noeqref{n_a_direct_sum_1_2_3-apx}
\noeqref{n_b_c_direct_sum_1_2_3-apx}
\noeqref{n_1_2_3_direct_sum_1_2_3-apx}
\noeqref{theorem_n_a-apx}
\noeqref{theorem_n_b_c-apx}
\noeqref{theorem_n_1_2_3-apx}

\title{Tripartite entanglement of qudits}

\author[1]{Roman V. Buniy\thanks{roman.buniy@gmail.com}}
\author[2]{Thomas W. Kephart\thanks{tom.kephart@gmail.com}}
\affil[1]{Schmid College of Science, Chapman University, Orange, CA 92866}
\affil[2]{Department of Physics and Astronomy, Vanderbilt University, Nashville, TN 37235}

\date{\today}

\maketitle

\begin{abstract}
    We provide an in-depth study of tripartite entanglement of qudits.
    We start with a short review of tripartite entanglement invariants, prove a theorem about the complete list of all allowed values of three (out of the total of four) such invariants, and give several bounds on the allowed values of the fourth invariant.
    After introducing several operations on entangled states (that allow us to build new states from old states) and deriving general properties pertaining to their invariants, we arrive at the decomposition theorem as one of our main results.
    The theorem relates the algebraic invariants of any entanglement class with the invariants of its corresponding components in each of its direct sum decompositions.
    This naturally leads to the definition of reducible and irreducible entanglement classes. 
    We explicitly compute algebraic invariants for several families of irreducible classes and show how the decomposition theorem allows computations of invariants for compounded classes to be carried out efficiently.
    This theorem also allows us to compute the invariants for the infinite number of entanglement classes constructed from irreducible components. 
    We proceed with the complete list of the entanglement classes for three tribits with decompositions of each class into irreducible components, and provide a visual guide to interrelations of these decompositions.
    We conclude with numerous examples of building classes for higher-spin qudits.
\end{abstract}

%------------------------------------------------------------------------------
\section{Introduction} 
%------------------------------------------------------------------------------

Many potential applications of quantum coherence and many aspects of quantum correlations have been reviewed in \cite{Horodecki:2009zz}.
It is not unlikely that quantum computing \cite{Steane:1997kb,PreskillLecture} will be the most important and widespread application, but it is far from obvious what the final form of the architecture of a quantum computer will be, or whether a single alternative will dominate.
What will be the quantum entities that are entangled: (magnetic) Josephson junctions, quantum dots, anyons, photons, atoms, etc.?
What spin should we choose: qubits, tribits, or higher-spin objects (qudits)?
Different choices have advantages and disadvantages and  choices will depend on a variety of compromises.
E.g., anyons are fault tolerant \cite{Kitaev:1997wr,Nayak:2008zza,NielsenChuang} but may be difficult to scale up.
Magnetic systems \cite{Castelvecchi} may be difficult to miniaturize and run at high clock speeds, etc.
All these  systems must be first understood at the fundamental level before one can confidently judge their potential.
Consequently, at this stage of development, we believe that it is appropriate to take a broad view and explore general questions of how systems can be entangled and how that information can be useful in informing later design choices.

We have started such a program in \cite{Buniy:2010yh,Buniy:2010zp}, where we introduced algebraic invariants for entangled quantum states and used values of these invariants to group states into equivalence classes describing their entanglement properties.
We applied this entanglement classification to three- and four-qubit systems as well as to numerous tripartite systems, for which we gave the full lists of all entanglement classes and representative states of each class (from which all other states in each class can be obtained by simple transformations of bases).
We also gave some general results for various qudit systems.
Here we extend our previous work by focusing on both general properties of three-qudit states and specific properties of three-tribit states.
We show how these two groups of states are interconnected by studying various methods of building new states from old states, which in particular help to construct infinite families of three-qudit states from three-tribit states.

Tripartite entanglement is particularly interesting as it opens up many new possibilities beyond simple bipartite entanglement, but it still presents a manageable set of 
options.
Quadripartite entanglement is already getting much more complicated to use and control and its full use is beyond any  practical applications that can be contemplated at present as we will comment on below. 

Qudits show promise for many reasons. In certain cases they may be able to speed up computations \cite{Neeley}; in some circumstances they can be easily manipulated, e.g., in diffractive optical devices \cite{Lima}. They have even been used to model microtubules in the brain \cite{Srivastava}. 
Once a full understanding of possible entanglements is in hand, then there is any number of potential uses of the results.
For instance, our previous work found application to the hierarchy of graph diagonal states \cite{Jafarizadeh} relating quantum to classical uncertainty \cite{Wang}.
Early studies of qudit entanglement can be found in \cite{Rungta} and \cite{Jamiolkowski}.
For a recent study of multi-qudit state generation see \cite{Ho:2016gki}.

Because of their intermediate position within the universe of entanglement, tripartite entangled states have been studied for over two decades and it is worthwhile to review a few highlights.
Tripartite entanglement provides a window into higher-order entanglement that is still sufficiently simple to allow us to gain valuable insights beyond what has been obtained from bipartite states.
Various aspects of tripartite entanglement have been considered over the previous decade that go into aspects of quantum fields and beyond, including string theory, supergravity and black holes.
For three qubits there is a relationship with the Cayley hyperdeterminant. This result has various generalizations to $N=2$, $4$ and $8$ supergravity, where the $N=8$ case corresponds to tripartite entanglement of seven qubits living in $E_7$ \cite{Duff:2006ue,Levay:2006pt}.
There is also a relationship between this result and entanglement entropy of STU black holes \cite{Borsten:2008wd,Borsten:2010db,Borsten:2012fx,Levay:2006kf}. 
The manipulation of entangled quantum states, including entanglement transfer, is all-important in applications; e.g., \cite{Bina,Kim:2015dbb}  have studied tripartite entanglement transfer in resonance cavities and  \cite{Hwang:2010ib,Shamirzai:2011gk,Khan:2014fna} have shown how entanglement degrades due to the Unruh effect in non-inertial frames.
It has also been shown how tripartite entanglement can be extracted from the vacuum \cite{Lorek}. 
Tripartite entanglement has played a major role in our understanding of a wide ranging set of physical situations, from the study of holography in black holes via von Neumann entropy \cite{Rota:2015wge}, to our understanding of quantum phase transitions \cite{Bayat:2017bnj}.

Interesting work toward practical application of qudit entanglement has appeared recently.
In particular, machine learning algorithms have been employed to find new quantum experiments \cite{Krenn2016,Melnikov2018} and described them in terms of Schmidt rank vectors \cite{Huber2013}.
It can be difficult for human intuition to arrive at such conclusions as they are the result of creating a toolbox of known experiments and then combining them in a multiplicity of ways until new useful experiments arise. 
The process can then be iterated with a new extended toolbox, etc.
The first level may be easy to grasp, but less so for successive levels, even though that is potentially where the practical applications lie.

Entangled qudits are also the basis of another fast-developing area of research: the entanglement of  multi-level photons \cite{Malik2016,Erhard2018,Babazadeh2017}. Devices based on entangled photons  hold the prospect of easy high-speed manipulation and interrogation and it would not be surprising if they become the key component in practical devices.

Hence we believe a comprehensive study of entangled tripartite qudits is in order, and that is what we undertake below.
We keep the development as general as possible by first studying properties of any tripartite entangled system that can be derived for any dimensionalities of the relevant Hilbert spaces.
We apply these results to particular cases of low dimensions: three qubits, two qubits and a tribit, one qubit and two tribits, and three tribits.
These examples serve both as elucidating examples of the general method, as well as a springboard into higher-dimensional qudits for which they play the role of their building blocks.

The rest of the article is organized as follows.
In Section~\ref{section_algebraic_invariants} we briefly review algebraic entanglement invariants for tripartite systems, where we limit ourselves to only the necessary notations and definitions.
In Section~\ref{section_general_results} we present general results on arbitrary systems of tripartite qudits (with some of them derived and the rest conjectured), which are exemplified by specific results in later sections.
In particular, we prove a theorem that gives the complete list of all allowed values for the first three (out of the total of four) algebraic invariants in any tripartite system.
It is convenient, and perhaps even required, to organize entangled states by expressing them (whenever possible) in terms of simpler states.
We achieve this by introducing operations that allow us to construct new entangled states from old states and by finding relations between their invariants.
Section~\ref{section_operations} is devoted to this task, whose development leads to the decomposition theorem, which is our main general result.
This theorem relates the algebraic invariants of any class with the invariants of its corresponding components in the direct sum decomposition, which helps with efficient calculations of entanglement invariants and the resulting organization of entangled classes into infinite families.
Consequently, in Section~\ref{section_decomposition} we introduce reducible and irreducible entanglement classes.
The next step is explicit computations of invariants for various irreducible classes, to which we turn in Section~\ref{section_irreducible_classes}.
We proceed by showing in Section~\ref{section_3_3_3} how these irreducible classes are sufficient to build all entanglement classes for three tribits.
In addition to the construction of these classes by means of tensor products and direct sums, we also demonstrate their destruction by annihilation operators, which provide complementary aspects of interrelations between various classes.
Using the decomposition theorem, we give in Section~\ref{section_beyond} numerous examples of building classes for higher-spin qudits and show that in general there are multiple ways to decompose a given entanglement class into irreducible components.
Finally, we conclude with a brief summary and outlook in Section~\ref{section_summary}, where we also list some specific unsolved problems within our restricted entanglement classification.

\section{Algebraic invariants for tripartite entanglement} \label{section_algebraic_invariants}

We start with a very brief review of the algebraic entanglement invariants for tripartite systems following the introduction of these invariants for any $n$-partite systems in \cite{Buniy:2010yh,Buniy:2010zp}.
We limit ourselves here to only the necessary notations and definitions.

We consider finite-dimensional vector spaces $\{V_a\}_{1\le a\le 3}$ and their tensor products $V_{b,c}=V_b\otimes V_c$, where $b,c\in\{1,2,3\}$, $b\not=c$, and $V_{a,b,c}=V_a\otimes V_b\otimes V_c$, where $\{a,b,c\}=\{1,2,3\}$.
(We use the standard convention that two equal sets $\{a,b,c\}$ and $\{1,2,3\}$ remain equal for any permutations of their elements.)
We take $\dim{V_a}=d_a$, in which case $\dim{V_{b,c}}=d_b d_c$, $\dim{V_{a,b,c}}=d_a d_b d_c$.

For each vector $v\in V_{1,2,3}$ and each $\{a,b,c\}=\{1,2,3\}$, we define the kernel spaces
\begin{align}
    &K_a(v)=\{ w\in V_a \colon C_a{(v\otimes w^*)}=0 \}, \label{k_a} \\
    &K_{b,c}(v)=\{ w\in V_{b,c} \colon C_b C_c(v\otimes w^*)=0 \}, \label{k_b_c} \\
    &K_{a,b,c}(v)=\{ w\in V_{a,b,c} \colon C_a C_b(v\otimes w^*)=0, \ C_a C_c(v\otimes w^*)=0, \ C_b C_c(v\otimes w^*)=0 \}. \label{k_a_b_c}
\end{align}
Here $w^*$ is the algebraic dual of $w$, $C_a$ is the contraction operator that acts on $V_a\otimes V^*_a$, and $V^*_a$ is the space that is algebraically dual to $V_a$. 
We look at the dimensions of these spaces, i.e., the nullities of the corresponding linear maps in \eqref{k_a}, \eqref{k_b_c} and \eqref{k_a_b_c},
\begin{align}
    &n_a(v)=\dim{K_a(v)}, \label{n_a} \\
    &n_{b,c}(v)=\dim{K_{b,c}(v)}, \label{n_b_c} \\
    &n_{a,b,c}(v)=\dim{K_{a,b,c}(v)}, \label{n_1_2_3} \noeqref{n_1_2_3}
\end{align}
and call them the algebraic entanglement invariants of $v$.

Since there are natural isomorphisms $V_{b,c}\simeq V_{c,b}$, $V_{a,b,c}\simeq V_{a',b',c'}$, $\{a',b',c'\}=\{a,b,c\}$, it is sufficient to consider only the tensor products $V_{1,2}, V_{1,3}, V_{2,3}, V_{1,2,3}$. 
Furthermore, it follows from their definitions \eqref{k_a} and \eqref{k_b_c} that the spaces in each pair $(K_a(v),K_{b,c}(v))$, where $\{a,b,c\}=\{1,2,3\}$, are simply related and consequently 
\begin{align}
    &d_a-n_a(v)=d_b d_c-n_{b,c}(v), \ \{a,b,c\}=\{1,2,3\}. \label{n_a_n_b_c}
\end{align}
As a result, for each $v\in V_{1,2,3}$, it is sufficient to consider only the kernel spaces $K_1(v)$, $K_2(v)$, $K_3(v)$, $K_{1,2,3}(v)$ and their dimensions $n_1(v)$, $n_2(v)$, $n_3(v)$, $n_{1,2,3}(v)$. 

For each vector $v\in V_{1,2,3}$, we define the entanglement class $C(v)$ associated with $v$ as the collection of all vectors $v'\in V_{1,2,3}$ that have the same algebraic invariants as $v$, so that
\begin{align}
    C(v)=\bigl\{ &v'\in V_{1,2,3} \colon n_a(v')=n_a(v), \ n_{b,c}(v')=n_{b,c}(v), \ n_{a,b,c}(v')=n_{a,b,c}(v), \{a,b,c\}=\{1,2,3\} \bigr\}
    \label{entanglement_class_c_v}
\end{align}
In view of \eqref{n_a_n_b_c}, each class $C(v)$ is fully determined by the values of only four invariants $n_1(v)$, $n_2(v)$, $n_3(v)$, $n_{1,2,3}(v)$.

For explicit computations of the algebraic entanglement invariants of $v$, we need expressions for the kernel spaces \eqref{k_a}, \eqref{k_b_c}, \eqref{k_a_b_c} in terms of coordinates.
We therefore choose $(e_{1,i})_{1\le i\le d_1}$, $(e_{2,j})_{1\le j\le d_2}$, $(e_{3,k})_{1\le k\le d_3}$ as respective bases for $V_1$, $V_2$, $V_3$.
The vectors $e_{1,i}\otimes e_{2,j}\otimes e_{3,k}$ make up the basis of $V_{1,2,3}$ and an arbitrary vector $v\in V_{1,2,3}$ can be written as
\begin{align}
    v&=\sum_{i=1}^{d_1} \sum_{j=1}^{d_2} \sum_{k=1}^{d_3} v_{i,j,k}e_{1,i}\otimes e_{2,j}\otimes e_{3,k}, \label{v_coordinates}
\end{align}
where $v_{i,j,k}$ are the coordinates of $v$ in the chosen basis of $V_{1,2,3}$.
The kernel spaces \eqref{k_a}, \eqref{k_b_c} and \eqref{k_a_b_c} become
\begin{align}
    &K_1(v)=\Bigl\{ w=\sum_{i=1}^{d_1} w_i e_{1,i}\in V_1 \colon \sum_{i=1}^{d_1} v_{i,j,k}w_i=0, \ 1\le j\le d_2, \ 1\le k\le d_3 \Bigr\}, \label{k_1_coordinates} \\
    &K_2(v)=\Bigl\{ w=\sum_{j=1}^{d_2} w_j e_{2,j}\in V_2 \colon \sum_{j=1}^{d_2} v_{i,j,k}w_j=0, \ 1\le i\le d_1, \ 1\le k\le d_3 \Bigr\}, \label{k_2_coordinates} \\
    &K_3(v)=\Bigl\{ w=\sum_{k=1}^{d_3} w_k e_{3,k}\in V_3 \colon \sum_{k=1}^{d_3} v_{i,j,k}w_k=0, \ 1\le i\le d_1, \ 1\le j\le d_2 \Bigr\}, \label{k_3_coordinates} \\
    &K_{1,2}(v)=\Bigl\{ w=\sum_{i=1}^{d_1} \sum_{j=1}^{d_2} w_{i,j} e_{1,i}\otimes e_{2,j}\in V_{1,2} \colon \sum_{i=1}^{d_1} \sum_{j=1}^{d_2} v_{i,j,k}w_{i,j}=0, \ 1\le k\le d_3 \Bigr\}, \label{k_1_2_coordinates} \\
    &K_{1,3}(v)=\Bigl\{ w=\sum_{i=1}^{d_1} \sum_{k=1}^{d_3} w_{i,k} e_{1,i}\otimes e_{3,k}\in V_{1,3} \colon \sum_{i=1}^{d_1} \sum_{k=1}^{d_3} v_{i,j,k}w_{i,k}=0, \ 1\le j\le d_2 \Bigr\}, \label{k_1_3_coordinates} \\
    &K_{2,3}(v)=\Bigl\{ w=\sum_{j=1}^{d_2} \sum_{k=1}^{d_3} w_{j,k} e_{2,j}\otimes e_{3,k}\in V_{2,3} \colon \sum_{j=1}^{d_2} \sum_{k=1}^{d_3} v_{i,j,k}w_{j,k}=0, \ 1\le i\le d_1 \Bigr\}, \label{k_2_3_coordinates} \\
    &K_{1,2,3}(v)=\Bigl\{ w=\sum_{i=1}^{d_1} \sum_{j=1}^{d_2} \sum_{k=1}^{d_3} w_{i,j,k}e_{1,i}\otimes e_{2,j}\otimes e_{3,k}\in V_{1,2,3} \colon \sum_{j=1}^{d_2} \sum_{k=1}^{d_3} v_{i,j,k}w_{i_0,j,k}=0, \ 1\le i,i_0\le d_1 ; \nn \\
    &\quad \sum_{i=1}^{d_1} \sum_{k=1}^{d_3} v_{i,j,k}w_{i,j_0,k}=0, \ 1\le j,j_0\le d_2; \ \sum_{i=1}^{d_1} \sum_{j=1}^{d_2} v_{i,j,k}w_{i,j,k_0}=0, \ 1\le k,k_0\le d_3 \Bigr \}. \label{k_1_2_3_coordinates}
\end{align}

Whenever it is convenient, we use the notation $[i,j,k]=e_{1,i}\otimes e_{2,j}\otimes e_{3,k}$.
We also often do not explicitly write $v$ in $K_a(v)$, $K_{b,c}(v)$$K_{1,2,3}(v)$, $n_a(v)$, $n_{b,c}(v)$, $n_{1,2,3}(v)$ if it is clear which $v$ is meant.

\section{General results on tripartite invariants} \label{section_general_results}

For entangled states with certain specific features, it is possible to find detailed relations (both inequalities and equalities) between their invariants.
We consider such cases in Section~\ref{section_operations}.
For now, however, we limit ourselves to the more general situations where states do not have any specific features, and, perhaps, it should not be surprising that all results of this kind are in the weaker form of inequalities. 
Some of our results in this section are conjectures for which we do not have proofs but instead have extensive empirical evidence supporting them.

In view of the significant differences between their definitions, it is much easier to establish general results for $n_1$, $n_2$, $n_3$ than for $n_{1,2,3}$.
We therefore first proceed with inequalities for $n_1,n_2,n_3$ when $n_{1,2,3}$ is not constrained, and then consider the problem of inequalities for $n_{1,2,3}$ when $n_1,n_2,n_3$ are constrained.

%------------------------------------------------------------------------------
\subsection{The invariants $\bm{n_1}$, $\bm{n_2}$, $\bm{n_3}$}
%------------------------------------------------------------------------------

%------------------------------------------------------------------------------
\begin{theoremrep}
    The inequalities
    \begin{align}
	&\max{(0,d_a-d_b d_c)}\le n_a\le d_a, \label{n_a_inequality} \\
	&d_a-n_a\le(d_b-n_b)(d_c-n_c) \label{d_a_n_a_inequality}
    \end{align}
    for all $\{a,b,c\}=\{1,2,3\}$ are necessary and sufficient conditions for the integers $n_1$, $n_2$, $n_3$ to be the first three nullities of any given $v\in V_{1,2,3}$.

    As a result, either $n_a=d_a$, $1\le a\le 3$ or
    \begin{align}
	d_3-(d_1-n_1)(d_2-n_2)\le n_3\le d_3-\max{\Bigl(\frac{d_1-n_1}{d_2-n_2},\frac{d_2-n_2}{d_1-n_1}\Bigr)} \label{n_3_inequality}
    \end{align}
    for $0\le n_a\le d_a-1$ and $d_1\le d_2\le d_3$.
    \label{theorem_general_n_a}
\end{theoremrep}
%------------------------------------------------------------------------------

\begin{toappendix}
    \label{appendix_theorem_general_n_a}
\end{toappendix}

See Appendix \ref{appendix_theorem_general_n_a} for the proof.

%------------------------------------------------------------------------------
\begin{appendixproof}
    We first prove the necessity.
    We take an arbitrary $v\in V_{1,2,3}$ and let $n_1(v)$, $n_2(v)$, $n_3(v)$ be its first three nullities. 
    It follows from the definitions \eqref{n_a} and \eqref{n_b_c} that the nullities $n_a$ and $n_{b,c}$ satisfy
    \begin{align}
	&0\le n_a(v)\le d_a, \\
	&0\le n_{b,c}(v)\le d_b d_c.
    \end{align}
    Using now \eqref{n_a_n_b_c}, we arrive at
    \begin{align}
	&\max{(0,d_a-d_b d_c)}\le n_a(v)\le d_a, \\
	&\max{(0,d_b d_c-d_a)}\le n_{b,c}(v)\le d_b d_c,
    \end{align}
    the first of which is \eqref{n_a_inequality-apx}.

    To prove \eqref{d_a_n_a_inequality-apx}, we use the fact that the space $V_{1,2,3}$ can be written in any of the three equivalent forms, $V_{1,2,3}=V_1\otimes V_{2,3}=V_2\otimes V_{1,3}=V_3\otimes V_{1,2}$.
    We start with the Schmidt decomposition for any given $v\in V_1\otimes V_{2,3}$, which states that there exist orthonormal sets of vectors $\{\alpha^{(l)}\}_{l=1}^m\subset V_1$ and $\{\tilde\alpha^{(l)}\}_{l=1}^m\subset V_{2,3}$ together with a set of non-negative numbers $\{\lambda^{(l)}\}_{l=1}^m$, where $m=\min{(d_1,d_2 d_3)}$, such that
    \begin{align}
	&v=\sum_{l=1}^m v^{(l)}, \\
	&v^{(l)}=\lambda^{(l)}\alpha^{(l)}\otimes\tilde\alpha^{(l)}.
    \end{align}
    Now \eqref{k_a} for $a=1$ becomes
    \begin{align}
	K_1(v)=\{w\in V_1\colon \lambda^{(l)}C_1(\alpha^{(l)}\otimes w^*)=0, \ 1\le l\le m\}.
    \end{align}
    The equation $ \lambda^{(l)}C_1(\alpha^{(l)}\otimes w^*)=0$ constrains $w$ only if $\lambda^{(l)}>0$, in which case there is exactly one constraint on $w$.
    Since these constraints on $w$ are generically independent from each other and $\dim{K_1(v)}=n_1(v)$, it follows that $d_1-n_1(v)$ numbers among $\{\lambda^{(l)}\}_{l=1}^m$ are positive.
    As the number of linearly independent elements of $V_{2,3}$ is $d_2 d_3$, there is a sufficient number of vectors in $\{\tilde{\alpha}^{(l)}\}_{l=1}^m$ only if $d_1-n_1(v)\le d_2 d_3$.
    This inequality holds in view of \eqref{n_a_inequality} for $a=1$.

    Without any loss of generality, by rearranging the order of elements in $\{\lambda^{(l)}\}_{l=1}^m$, we choose all elements in $\{\lambda^{(l)}\}_{l=1}^{d_1-n_1(v)}$ to be positive and all elements in $\{\lambda^{(l)}\}_{l=d_1-n_1(v)+1}^m$ to be zero, and arrive at
    \begin{align*}
	&v=\sum_{l=1}^{d_1-n_1(v)} v^{(l)}, \\
	&v^{(l)}=\lambda^{(l)}\alpha^{(l)}\otimes\tilde\alpha^{(l)}.
    \end{align*}
    Now \eqref{k_a} for $a\in\{2,3\}$ become
    \begin{align}
	&K_a(v)=\cap_{l=1}^{d_1-n_1(v)}K_a(v^{(l)}), \label{k_a_schmidt} \\
	&K_a(v^{(l)})=\{w\in V_a\colon C_a(\tilde\alpha^{(l)}\otimes w^*)=0\}. \label{k_a_schmidt_component}
    \end{align}
    It is clear that
    \begin{align}
	&n_a(v^{(l)})=\dim{K_a(v^{(l)})}=d_a-\rank{\tilde{A}^{(l)}},
	\label{n_a_v_l}
    \end{align}
    where $\tilde{A}^{(l)}$ is the $d_2\times d_3$ matrix with the matrix elements $\tilde\alpha^{(l)}_{j,k}$ such that
    \begin{align*}
	\tilde\alpha^{(l)}=\sum_{j=1}^{d_2} \sum_{k=1}^{d_3} \tilde\alpha^{(l)}_{j,k}e_{2,j}\otimes e_{3,k},
    \end{align*}
    and $\rank{\tilde{A}^{(l)}}$ is the matrix rank of $\tilde{A}^{(l)}$.
    Comparing \eqref{n_a_v_l} for $a=2$ and $a=3$, we find
    \begin{align}
	d_2-n_2(v^{(l)})=d_3-n_3(v^{(l)}),
	\label{}
    \end{align}
    which expresses the equality of the row rank and column rank of $\tilde{A}^{(l)}$. 

    To find bounds on $d_a-n_a(v)=\dim{(V_a\setminus K_a(v))}$, we use de Morgan's laws to relate the complements of $K_a(v)$ and $\{K_a(v^{(l)})\}_{l=1}^{d_1-n_1(v)}$ in $V_a$,
    \begin{align}
	&V_a\setminus K_a(v)=\cup_{l=1}^{d_1-n_1(v)}\bigl(V_a\setminus K_a(v^{(l)})\bigr).
	\label{morgan}
    \end{align}
    The inclusion-exclusion principle allows us to express the dimension of the union of any set of spaces $\{S_l\}_{l=1}^p$ in terms of the dimensions of their intersections,
    \begin{align}
	\dim{\bigl(\cup_{l=1}^p S_l\bigr)} &=\sum_{l=1}^p \dim{S_l} -\sum_{1\le l_1<l_2\le p} \dim{(S_{l_1}\cap S_{l_2})} +\sum_{1\le l_1<l_2<l_3\le p} \dim{(S_{l_1}\cap S_{l_2}\cap S_{l_3})} \nn \\
	&-\dotsb +(-1)^{p-1} \dim{\bigl(S_1\cap\dotsb\cap S_p\bigr)}.
	\label{inclusion_exclusion}
    \end{align}
    If we do not know the dimensions of intersections of the spaces $\{S_l\}_{l=1}^p$, the best bounds that we can establish for $\dim{\bigl(\cup_{l=1}^p S_l\bigr)}$ in \eqref{inclusion_exclusion} are
    \begin{align}
	\max{\bigl\{\dim{S_l}\}_{l=1}^p}\le\dim{\bigl(\cup_{l=1}^p S_l\bigr)}\le\sum_{l=1}^p \dim{S_l}.
	\label{inclusion_exclusion_bounds}
    \end{align}
    Applying \eqref{inclusion_exclusion_bounds} to \eqref{morgan} and using \eqref{n_a} and \eqref{n_a_v_l}, we find
    \begin{align}
	\max{\bigl\{\rank{\tilde{A}^{(l)}}\bigr\}_{l=1}^{d_1-n_1(v)}}\le d_a-n_a(v)\le\sum_{l=1}^{d_1-n_1(v)} \rank{\tilde{A}^{(l)}}.
	\label{n_a_rank_A_bounds}
    \end{align}
    Since $\tilde{A}^{(l)}$ is a $d_2\times d_3$ matrix, we have $\rank{\tilde{A}^{(l)}}\le\min{(d_2,d_3)}$ and the lower bound in \eqref{n_a_rank_A_bounds} gives
    \begin{align}
	d_a-n_a(v) \ge\max{\bigl\{d_b-n_b(v^{(l)})\bigr\}_{l=1}^{d_1-n_1(v)}}, 
	\label{d_a_n_a_lower_bound}
    \end{align}
    or, equivalently,
    \begin{align}
	n_a(v) \le d_a-d_b+\min{\bigl\{n_b(v^{(l)})\bigr\}_{l=1}^{d_1-n_1(v)}}, 
	\label{}
    \end{align}
    both for any $a,b\in\{2,3\}$.
    The upper bound in \eqref{n_a_rank_A_bounds} becomes
    \begin{align}
	d_a-n_a(v) \le\sum_{l=1}^{d_1-n_1(v)} (d_b-n_b(v)) \le\sum_{l=1}^{d_1-n_1(v)} (d_b-n_b(v^{(l)})) =(d_1-n_1(v))(d_b-n_b(v))
	\label{d_a_n_a_upper_bound}
    \end{align}
    for any $\{a,b\}\subset\{2,3\}$.
    In the first inequality here we used \eqref{n_a_v_l}, and in the second inequality we used $d_b-n_b(v^{(l)})\le d_b-n_b(v)$ that follows from \eqref{d_a_n_a_lower_bound}.
    For $a=b$, \eqref{d_a_n_a_upper_bound} holds trivially since $d_1-n_1(v)\ge 1$. 
    For $a\not=b$, \eqref{d_a_n_a_upper_bound} is non-trivial and it becomes \eqref{d_a_n_a_inequality} for $(a,b)\in\{(2,3),(3,2)\}$.

    The remaining inequality for $a=1$ is similarly derived by using the Schmidt decomposition either for each $v\in V_2\otimes V_{1,3}$ or for each $v\in V_3\otimes V_{1,2}$, which completes the proof of \eqref{d_a_n_a_inequality} and the necessity.

    We now prove the sufficiency.
    Let $n_1$, $n_2$, $n_3$ be integers that satisfy \eqref{n_a_inequality} and \eqref{d_a_n_a_inequality} and let us assume, without any loss of generality, that $d_1-n_1\ge d_2-n_2\ge d_3-n_3$.
    We choose subsets $J_1,J_2,J_3$ such that $J_a\subset I_a$, where $I_a=\{1,\dotsc,d_a\}$, $\smallabs{J_a}=d_a-n_a$, $1\le a\le 3$.
    Let $\tilde{v}_3$ be a $(d_1-n_1)\times(d_2-n_2)$ matrix whose rows and columns are labeled by all the elements of $J_1$ and $J_2$, respectively, and which satisfy the following two conditions.
    First, there are exactly $d_3-n_3$ non-zero matrix elements of $\tilde{v}_3$, and the set of all their values coincides with $J_3$.
    This is always possible in view of the inequality \eqref{d_a_n_a_inequality} for $a=3$.
    Second, there is at most only one non-zero matrix element in each row and each column of $\tilde{v}_3$.  
    This is guaranteed by the inequalities $d_1-n_1\ge d_2-n_2\ge d_3-n_3$.
    All the remaining $(d_1-n_1)(d_2-n_2)-(d_3-n_3)$ matrix elements of $\tilde{v}_3$ equal zero.

    Let $v_3$ be the $d_1\times d_2$ matrix whose rows and columns are labeled by all the elements of $I_1$ and $I_2$, respectively, and such that $\tilde{v}_3$ is a submatrix of $v_3$ and all matrix elements of $v_3$ outside of $\tilde{v}_3$ are zero.
    Finally, we define a vector $v\in V_{1,2,3}$ in terms of the matrix elements $(v_3)_{i,j}$ of $v_3$ according to
    \begin{align*}
	v_{i,j,k}=
    %\delta_{(v_3)_{i,j},k}=
	\begin{cases}
	    1, & (v_3)_{i,j}=k, \\
	    0, & (v_3)_{i,j}=0.
	\end{cases}
    \end{align*}
    It follows that
    \begin{align}
	K_a(v)=\{w\in V_a\colon w_i=0, \ i\in J_a\}, \ 1\le a\le 3,
	\label{}
    \end{align}
    which implies $n_a(v)=n_a$, $1\le a\le 3$.
    This completes the proof of the sufficiency because we constructed a vector $v$ for which the nullities $n_a(v)$ are the given integer solutions of \eqref{n_a_inequality} and \eqref{d_a_n_a_inequality}.

    Finally, to prove \eqref{n_3_inequality-apx} we first note that it is easy to see from \eqref{d_a_n_a_inequality} that if any one quantity among $n_1, n_2, n_3$ takes its greatest value ($d_1, d_2, d_3$, respectively), then so do all other quantities.
    Setting aside this simple case of the maximal nullities (which corresponds to the entanglement class $C_0$, the class of the vacuum state, as shown below), we proceed with the remaining case $0\le n_a\le d_a-1$, $1\le a\le 3$.
    It is convenient to assume without any loss of generality that $1\le d_1\le d_2\le d_3$.
    Inequalities \eqref{n_a_inequality} for $a\in\{1,2\}$ are trivially satisfied.
    Solving the three inequalities \eqref{d_a_n_a_inequality}, $a\in\{1,2,3\}$ for $n_3$, we find \eqref{n_3_inequality}.
    Specifically, \eqref{d_a_n_a_inequality} for $a=3$ gives the first inequality in \eqref{n_3_inequality}, and \eqref{d_a_n_a_inequality} for $a=1$ and $a=2$ give the second inequality in \eqref{n_3_inequality}.
    Now it follows from the first inequality in \eqref{n_3_inequality} that \eqref{n_a_inequality} for $a=3$ is trivially satisfied because $d_1 d_2\ge (d_1-n_1)(d_2-n_2)$.
    As a result, the only and all possible values of $n_1, n_2, n_3$ are either $n_1=d_1$, $n_2=d_2$, $n_3=d_3$ or the integers $n_1, n_2$ satisfying $0\le n_1\le d_1-1$, $0\le n_2\le d_2-1$ and the integers $n_3$ satisfying \eqref{n_3_inequality}.

\end{appendixproof}
%------------------------------------------------------------------------------

%------------------------------------------------------------------------------
\subsubsection{The number of allowed values of $(n_1,n_2,n_3)$}
%------------------------------------------------------------------------------

Let $N_{d_1,d_2,d_3}$ denote the number of all possible values of nullities $(n_1,n_2,n_3)$ allowed for a given set of dimensions $(d_1,d_2,d_3)$.
Writing \eqref{n_a_inequality} and \eqref{d_a_n_a_inequality} in terms of the ranks $r_a=d_a-n_a$, $1\le a\le 3$ in the form
\begin{align}
    &0\le r_a\le\min{(d_a,d_b d_c)}, \\
    &r_a\le r_b r_c,
    \label{}
\end{align}
it follows from Theorem \ref{theorem_general_n_a} that
\begin{align}
    &N_{d_1,d_2,d_3} =\sum_{r_1=0}^{d_1} \sum_{r_2=0}^{d_2} \sum_{r_3=0}^{d_3} \Theta{(r_1,r_2,r_3)}, \label{n_d_1_d_2_d_3} \\
    &\Theta{(r_1,r_2,r_3)} =\theta{(r_1 r_2-r_3)} \theta{(r_1 r_3-r_2)} \theta{(r_2 r_3-r_1)}. \label{theta}
\end{align}
Here $\theta$ is the Heaviside step function, which is defined by $\theta{(x)=1}$ for $x\ge 0$ and $\theta{(x)}=0$ for $x<0$.
Since $\Theta{(r_1,r_2,r_3)}$ is symmetric with respect to permutations of $r_1$, $r_2$, $r_3$, it follows that $N_{d_1,d_2,d_3}$ is symmetric with respect to permutations of $d_1$, $d_2$, $d_3$.

It appears that \eqref{n_d_1_d_2_d_3} is an interesting mathematical quantity that is worthwhile computing.
Our interest in it lies in the fact that the number of entanglement classes for $(d_1,d_2,d_3)$ is greater or equal to $N_{d_1,d_2,d_3}$.
This follows immediately from the fact that for each allowed set of values of $(n_1,n_2,n_3)$ there is at least one allowed value of $n_{1,2,3}$.

As we cannot compute \eqref{n_d_1_d_2_d_3} for arbitrary $(d_1,d_2,d_3)$, we look at exact values of $N_{d_1,d_2,d_3}$ only for several specific cases.
For the general case we content ourselves with establishing asymptotic expressions for $N_{d_1,d_2,d_3}$ for large dimensions.

Writing \eqref{n_d_1_d_2_d_3} in the form
\begin{align}
    &N_{d_1,d_2,d_3} =\Theta{(0,0,0)} +\sum_{r_1=1}^{d_1} \Theta{(r_1,0,0)} +\sum_{r_2=1}^{d_2} \Theta{(0,r_2,0)} +\sum_{r_3=1}^{d_3} \Theta{(0,0,r_3)} \nn \\
    &+\sum_{r_1=1}^{d_1} \sum_{r_2=1}^{d_2} \Theta{(r_1,r_2,0)}
    +\sum_{r_1=1}^{d_1} \sum_{r_3=1}^{d_3} \Theta{(r_1,0,r_3)}
    +\sum_{r_2=1}^{d_2} \sum_{r_3=1}^{d_3} \Theta{(0,r_2,r_3)}
    +\sum_{r_1=1}^{d_1} \sum_{r_2=1}^{d_2} \sum_{r_3=1}^{d_3} \Theta{(r_1,r_2,r_3)}
    \label{}
\end{align}
and using $\Theta{(0,0,0)}=1$ and $\Theta{(r_1,r_2,r_3)}=0$ if there are exactly one or two zeroes among $r_1$, $r_2$, $r_3$, we find
\begin{align}
    &N_{d_1,d_2,d_3} =1 +\sum_{r_1=1}^{d_1} \sum_{r_2=1}^{d_2} \sum_{r_3=1}^{d_3} \Theta{(r_1,r_2,r_3)}.
    \label{n_d_1_d_2_d_3_all_summations_start_at_1}
\end{align}
If at least one of the dimensions $d_1,d_2,d_3$ equals unity, then only the term $\Theta{(1,1,1)}=1$ contributes to \eqref{n_d_1_d_2_d_3_all_summations_start_at_1}, which results in 
\begin{align}
    N_{1,1,1} =N_{d_1,1,1} =N_{1,d_2,1} =N_{1,1,d_3} =N_{d_1,d_2,1} =N_{d_1,1,d_3} =N_{1,d_2,d_3} =2
    \label{}
\end{align}
for any $d_1\ge 1$, $d_2\ge 1$, $d_3\ge 1$.

It is easy to establish simple lower and upper bounds for $N_{d_1,d_2,d_3}$ for arbitrary $d_1,d_2,d_3$.
Indeed, since $\Theta$ satisfies $0\le\Theta(r_1,r_2,r_3)\le 1$ for any $r_a$ satisfying $1\le r_a\le d_a$ for $1\le a\le 3$, we conclude that
\begin{align}
    1\le N_{d_1,d_2,d_3}\le d_1 d_2 d_3 +1.
    \label{}
\end{align}

\begin{figure}[htpb]
    \centering
    \includegraphics[width=0.35\textwidth]{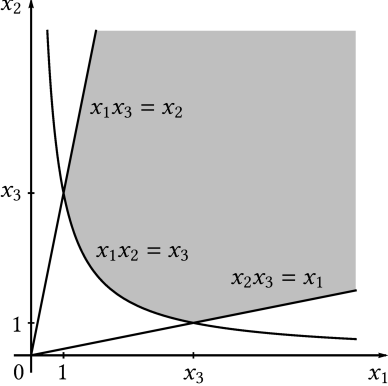}
    \caption{The shaded region is the part of the integration region for the variables $x_1$ and $x_2$ that gives the non-zero contribution to the integral in \eqref{n_d_1_d_2_d_3_integral} for any fixed $1<x_3$ for the case $d_1=d_2=d_3=d$.
    The displayed equations of two lines and a branch of a hyperbola follow from the definition of the function $\Theta(x_1,x_2,x_3)$ in \eqref{n_d_1_d_2_d_3_integral}.}
    \label{figure_integration_region}
\end{figure}

We can write the triple sum in  \eqref{n_d_1_d_2_d_3_all_summations_start_at_1} as the Riemann sum with the midpoint rule of the corresponding integral, which leads to the approximation
\begin{align}
    &N_{d_1,d_2,d_3} \approx \tilde{N}_{d_1,d_2,d_3}, \\
    &\tilde{N}_{d_1,d_2,d_3}=1+\int_{\frac{1}{2}}^{d_1+\frac{1}{2}} \int_{\frac{1}{2}}^{d_2+\frac{1}{2}} \int_{\frac{1}{2}}^{d_3+\frac{1}{2}} \Theta{(x_1,x_2,x_3)}\, dx_1\, dx_2\, dx_3
    \label{n_d_1_d_2_d_3_integral}
\end{align}
that becomes exact for $d_1,d_2,d_3\to\infty$.
Even though the integral in \eqref{n_d_1_d_2_d_3_integral} can be computed analytically for arbitrary $d_1$, $d_2$, $d_3$, the result is somewhat lengthy and we give the answer only for the case $d_1=d_2=d_3=d$.
We first compute the double integral over $x_1$ and $x_2$, which equals the area of the shaded region on Figure \ref{figure_integration_region}.

%------------------------------------------------------------------------------
%------------------------------------------------------------------------------

This area is $0$ for $x_3\le 1$ and
\begin{align}
    &\int_{\frac{x_3}{d+\frac{1}{2}}}^{d+\frac{1}{2}} \bigl(d+\tfrac{1}{2}-x_3 x_1^{-1}\bigr) \, dx_1 -2\int_{x_3}^{d+\frac{1}{2}} \bigl(x_1 x_3^{-1}-x_3 x_1^{-1}\bigr) \, dx_1 =(d+\tfrac{1}{2})^2(1-x_3^{-1}) -x_3\ln{x_3}
    \label{}
\end{align}
for $x_3>1$.
Integrating this expression over $x_3\in[1,d+\frac{1}{2}]$, we find
\begin{align}
    \tilde{N}_{d,d,d}=d^3+\tfrac{3}{4}d^2+\tfrac{11}{16} -\tfrac{3}{2}(d+\tfrac{1}{2})^2\ln{(d+\tfrac{1}{2})}.
    \label{}
\end{align}
The quantity $\tilde{N}_{d,d,d}$ approaches $d^3$ asymptotically.
In Table \ref{table_number_of_n_1_n_2_n_3_for_large_dimensions} we give the exact values $N_{d,d,d}$, the approximate values $\tilde{N}_{d,d,d}$ (rounded to the nearest integers) and the relative errors $(\tilde{N}_{d,d,d}/N_{d,d,d})-1$ for several values of $d$.

%------------------------------------------------------------------------------
\begin{table}[htpb]
    \centering
    \begin{tabular}{cccc}
	\toprule
	$d$ & $N_{d,d,d}$ & $\tilde{N}_{d,d,d}$ & $(\tilde{N}_{d,d,d}/N_{d,d,d})-1$ \\ \midrule
	$1$ & $2$ & $1$ & $-0.4655$ \\
	$2$ & $6$ & $3$ & $-0.4838$ \\
	$5$ & $75$ & $67$ & $-0.1055$ \\
	$10$ & $701$ & $687$ & $-0.0202$ \\
	$20$ & $6399$ & $6397$ & $-0.0004$ \\
	$50$ & $111591$ & $111873$ & $0.0025$ \\
	$100$ & $936038$ & $937655$ & $0.0017$ \\
	$200$ & $7702692$ & $7710360$ & $0.0010$ \\
	$500$ & $122798730$ & $122851984$ & $0.0004$ \\
	\bottomrule
    \end{tabular}
    \caption{The quantities $N_{d,d,d}$, their approximations $\tilde{N}_{d,d,d}$ (rounded to the nearest integers) and the relative errors $(\tilde{N}_{d,d,d}/N_{d,d,d})-1$ (rounded to the nearest $0.0001$) for several values of $d$.
    }
    \label{table_number_of_n_1_n_2_n_3_for_large_dimensions}
\end{table}
%------------------------------------------------------------------------------

As a result, we established an approximate lower bound for the number of entanglement classes,
\begin{align}
    \#\{C(v)\colon v\in V_{1,2,3}\}\ge N_{d_1,d_2,d_3}\approx \tilde{N}_{d_1,d_2,d_3}.
    \label{}
\end{align}

%------------------------------------------------------------------------------
\subsection{The invariant $\bm{n_{1,2,3}}$}
%------------------------------------------------------------------------------

We now turn to the nullity $n_{1,2,3}$.
From the definition
\begin{align}
    K_{1,2,3}(v) &= \ker{(f_{1,2}(v)\otimes\id_3)}\cap\ker{(f_{1,3}(v)\otimes\id_2)}\cap\ker{(f_{2,3}(v)\otimes\id_1)}, 
    \label{}
\end{align} 
where $f_{b,c}(v)$ is the linear map for which $K_{b,c}(v)$ is the kernel space, we find
\begin{align}
    &\dim{K_{1,2,3}(v)}\le\dim{(\ker{f_{1,2}(v)})}\dim{V_3},\\
    &\dim{K_{1,2,3}(v)}\le\dim{(\ker{f_{1,3}(v)})}\dim{V_2},\\
    &\dim{K_{1,2,3}(v)}\le\dim{(\ker{f_{2,3}(v)})}\dim{V_1},
    \label{}
\end{align}
which together with \eqref{n_a_n_b_c} imply
\begin{align}
    n_{1,2,3}\le d_1 d_2 d_3-\max\bigl\{d_1(d_1-n_1), d_2(d_2-n_2), d_3(d_3-n_3)\bigr\}. \label{n_1_2_3_upper_bound}
\end{align}

We can derive an upper bound similar to the one in \eqref{n_1_2_3_upper_bound} with the help of the method used in the proof of Theorem~\ref{theorem_general_n_a}.
Using one of the vectors $v$ constructed there for which the nullities $n_a(v)$ are equal to given integer solutions of \eqref{n_a_inequality} and \eqref{d_a_n_a_inequality} (and remembering the assumption $d_1-n_1\le d_2-n_2\le d_3-n_3$), we find that the system of equations $\sum_{i,j} v_{i,j,k}w_{i,j,k_0}=0$, $1\le k,k_0\le d_3$ in the definition \eqref{k_1_2_3_coordinates} of $K_{1,2,3}(v)$ becomes $\{w_{i,j,k_0}=0\colon v_{i,j,k}=0, \ 1\le k,k_0\le d_3\}$.
As it is harder to simplify the remaining two sets of equations in \eqref{k_1_2_3_coordinates}, we content ourselves with a weaker bound by considering only one set of the constraints in \eqref{k_1_2_3_coordinates}, use $d_3-n_3\le (d_1-n_1)(d_2-n_2)$ and conclude that
\begin{align}
    n_{1,2,3}\le d_1 d_2 d_3-d_3(d_3-n_3) \ \textrm{\ when \ } d_1-n_1\le d_2-n_2\le d_3-n_3. \label{}
\end{align}

We have extensive empirical evidence supporting the following conjectures for lower bounds of $n_{1,2,3}$.

%------------------------------------------------------------------------------
\begin{conjecture}
    The nullity $n_{1,2,3}$ satisfies
    \begin{align}
	&n_{1,2,3}\ge\max{\bigl\{(d_{1} d_{2}-d_{3}+n_{3})n_{3},(d_{1} d_{3}-d_{2}+n_{2})n_{2},(d_{2} d_{3}-d_{1}+n_{1})n_{1}\bigr\}}. \label{n_1_2_3_conjecture_2}
    \end{align}
    \label{conjecture_2}
\end{conjecture}
%------------------------------------------------------------------------------

%------------------------------------------------------------------------------
\begin{conjecture}
    The nullity $n_{1,2,3}$ satisfies
    \begin{align}
	n_{1,2,3}&\ge n_{1}\bigl((d_{2}-n_{2})(d_{3}-n_{3})-(d_{1}-n_{1})\bigr)+n_{2}\bigl((d_{1}-n_{1})(d_{3}-n_{3})-(d_{2}-n_{2})\bigr)\nn\\
	&+n_{3}\bigl((d_{1}-n_{1})(d_{2}-n_{2})-(d_{3}-n_{3})\bigr)\nn\\
	&+\max{\bigl\{n_{1}(d_{2} n_{3}+n_{2} d_{3}-n_{2} n_{3}), n_{2}(d_{1} n_{3}+n_{1} d_{3}-n_{1} n_{3}), n_{3}(d_{1} n_{2}+n_{1} d_{2}-n_{1} n_{2})\bigr\}}. \label{n_1_2_3_conjecture_3}
    \end{align}
    \label{conjecture_3}
\end{conjecture}
%------------------------------------------------------------------------------

%------------------------------------------------------------------------------
\begin{conjecture}
    The nullity $n_{1,2,3}$ satisfies
    \begin{align}
	n_{1,2,3}&\ge n_{1}\bigl((d_{2}-n_{2})(d_{3}-n_{3})-(d_{1}-n_{1})\bigr)+n_{2}\bigl((d_{1}-n_{1})(d_{3}-n_{3})-(d_{2}-n_{2})\bigr)\nn\\
	&+n_{3}\bigl((d_{1}-n_{1})(d_{2}-n_{2})-(d_{3}-n_{3})\bigr)\nn\\
	&+\max{\bigl\{n_{1}(d_{2} n_{3}+n_{2} d_{3}-n_{2} n_{3}), n_{2}(d_{1} n_{3}+n_{1} d_{3}-n_{1} n_{3}), n_{3}(d_{1} n_{2}+n_{1} d_{2}-n_{1} n_{2})\bigr\}}\nn\\
	&+\min{\bigl\{(d_{1}-n_{1})n_{2} n_{3},n_{1}(d_{2}-n_{2})n_{3},n_{1} n_{2}(d_{3}-n_{3})\bigr\}}. \label{n_1_2_3_conjecture_4}
    \end{align}
    \label{conjecture_4}
\end{conjecture}
%------------------------------------------------------------------------------

%------------------------------------------------------------------------------
\section{Operations on entangled states} \label{section_operations}
%------------------------------------------------------------------------------

We now study several operations that allow us to build new states from old states.
We restrict our attention to operations that take two vectors $v'$ and $v''$ as an input and produce a vector $v$ as an output.
More complicated new vectors can be constructed by repeated use of such operations.
Our goal is to find out how the invariants of $v$ are related to the invariants of $v'$ and $v''$.
For most of the operations that we consider, only inequalities between these invariants can be derived.
For one operation, however, we find that the invariants of $v$ are determined by the invariants of $v'$ and $v''$.
These results can be used to bound or determine precisely invariants of more complicated vectors from their less complicated building blocks.
We now proceed with the description of several types of such operations.

%------------------------------------------------------------------------------
\subsection{Tensor products}
%------------------------------------------------------------------------------

The only non-trivial independent cases where the space $V=V_{1,2,3}$ can be written as the tensor product $V=V'\otimes V''$ are those where $(V',V'')\in\{(V_1,V_{2,3}),(V_2,V_{1,3}),(V_3,V_{1,2})\}$.
Let $(V',V'')=(V_1,V_2\otimes V_3)$.
For $v'\in V'$, $v''\in V''$, $v=v'\otimes v''\in V$, equations \eqref{k_a}, \eqref{k_b_c} and \eqref{k_a_b_c} become
\begin{align}
    &K_{1}(v'\otimes v'')=\{ w\in V_{1} \colon C_{1}(v'\otimes w^*)=0 \}, \\
    &K_{a}(v'\otimes v'')=\{ w\in V_{a} \colon C_{a}(v''\otimes w^*)=0 \}, \ 2\le a\le 3, \\
    &K_{1,a}(v'\otimes v'')=\{ w\in V_{1,a} \colon C_{1} C_{a}(v'\otimes v''\otimes w^*)=0 \}, \ 2\le a\le 3, \\
    &K_{2,3}(v'\otimes v'')=\{ w\in V_{2,3} \colon C_{2} C_{3}(v''\otimes w^*)=0 \}, \\
    &K_{1,2,3}(v'\otimes v'')=\{ w\in V_{1,2,3} \colon C_{1} C_{2}(v'\otimes v''\otimes w^*)=0, \ C_{1} C_{3}(v'\otimes v''\otimes w^*)=0, \ C_{2} C_{3}(v''\otimes w^*)=0 \},
\end{align}
which lead to
\begin{align}
    &n_{1}(v'\otimes v'')=d_{1}-\rank{v'}, \\
    &n_{a}(v'\otimes v'')=d_{a}-\rank{v''}, \ 2\le a\le 3, \\
    &n_{1,a}(v'\otimes v'')=d_{1}d_{a}-\rank{v''}, \ 2\le a\le 3, \\
    &n_{2,3}(v'\otimes v'')=d_{2}d_{3}-\rank{v'}, \\
    &n_{1,2,3}(v'\otimes v'')\ge d_{1} d_{2} d_{3}-d_{1}-(d_{2}+d_{3})\rank{v''},
    \label{}
\end{align}
where $\rank{v'}=1$ and $\rank{v''}$ are the ranks of the $d_1\times 1$ matrix $v'$ and the $d_2\times d_3$ matrix $v''$, respectively. 

The results for the remaining two cases $(V',V'')\in\{(V_2,V_{1,3}),(V_3,V_{1,2})\}$ follow from the above equations by permutation of indices.

%------------------------------------------------------------------------------
\subsection{Direct sums} \label{section_direct_sums}
%------------------------------------------------------------------------------

\begin{figure}[htpb]
    \centering
    \includegraphics[width=200pt]{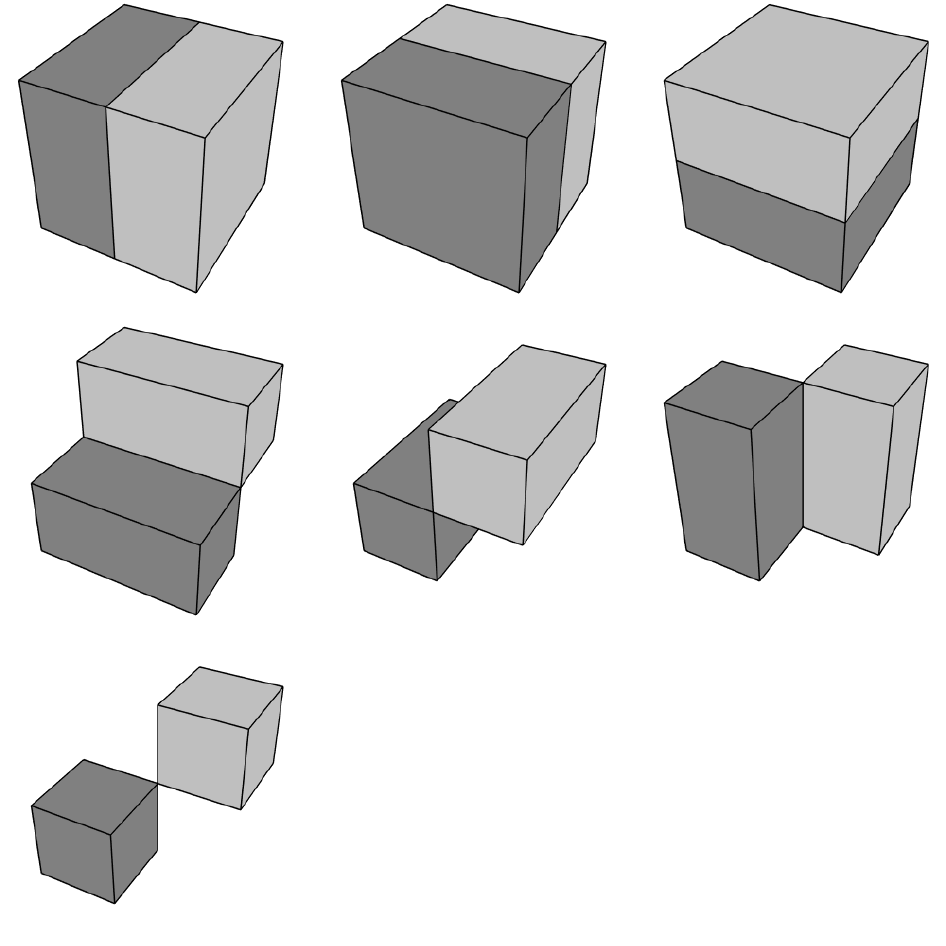}
    \caption{An illustration of how partitions of $S_1$, $S_2$, $S_3$ lead to diagonal subsets of the set $S=S_1\times S_2\times S_3$.
    In the first, second and third row, we partition one, two and three sets, respectively.}
    \label{figure_direct_sum_1_2_3}
\end{figure}

Let $1\le m\le 3$ and consider one of the $\binom{3}{m}$ choices to select $m$ sets among the sets $S_a=\{1,\dotsc,d_a\}$, $1\le a\le 3$.
We partition each of the selected $m$ sets $S_a$ into two subsets $S'_a$ and $S''_a$ and find the resulting partitioning of the set $S=S_1\times S_2\times S_3$ into $2^m$ subsets.
Among these $2^m$ subsets, we select two diagonal subsets $S'$ and $S''$ given by
\begin{align}
    &S'=
    \begin{cases}
	S'_1\times S_2\times S_3, & m=1, \\
	S'_1\times S'_2\times S_3, & m=2, \\
	S'_1\times S'_2\times S'_3, & m=3,
    \end{cases}
    \\
    &S''=
    \begin{cases}
	S''_1\times S_2\times S_3, & m=1, \\
	S''_1\times S''_2\times S_3, & m=2, \\
	S''_1\times S''_2\times S''_3, & m=3
    \end{cases}
    \label{}
\end{align}
and the two other expressions for each $m=1$ and $m=2$ obtained by permutation of indices.
(Although the diagonal subsets for $m=1$ are the only subsets of $S$, we use this terminology for uniformity of notations.)
See Figure~\ref{figure_direct_sum_1_2_3} for an illustration.

%------------------------------------------------------------------------------
%------------------------------------------------------------------------------

The above partitions lead to the direct sum decompositions of each space $V_a$ such that $V_a=V'_a\oplus_a V''_a$, where $V'_a=\linearspan{\{e_{a,i}\}_{i\in S'_a}}$ and $V''_a=\linearspan{\{e_{a,i}\}_{i\in S''_a}}$.
We attach labels to the direct sum symbols ($\oplus_{a}$, $\oplus_{a,b}$, $\oplus_{1,2,3}$) to specify unambiguously the directions in which the direct sums act.
We similarly attach labels to the sum symbols ($+_{a}$, $+_{a,b}$, $+_{1,2,3}$) when adding vectors to specify unambiguously the directions in which the sums act.
We now consider each case $1\le m\le 3$ separately. 

In deriving several inequalities in the following lemmas, we use the following elementary fact about the numbers of constraints $N$ associated with arbitrary linear functions $\{F_i(w)\}_{i\in I}$ that are imposed on a vector $w\in W$:
\begin{align}
    &N\bigl[w\in W\colon \sum_{i\in I}F_i(w)=0\bigr] \le N\bigl[w\in W\colon \sum_{i\in J}F_i(w)=0,\sum_{i\in I\setminus J}F_i(w)=0\bigr] \nn \\
    &\quad \le\min\biggl\{\dim{W},\sum_{i\in J} N\bigl[w\in W\colon F_i(w)=0\bigr]+\sum_{i\in I\setminus J}N\bigl[w\in W\colon F_i(w)=0\bigr]\biggr\}
    \label{inequalities_for_systems_of_linear_equations}
\end{align}
for any $J\subset I$.

%------------------------------------------------------------------------------
\subsubsection{One decomposition} \label{section_direct_sum_1}
%------------------------------------------------------------------------------

%------------------------------------------------------------------------------
\begin{lemmarep}
    For a direct sum decomposition generated by $V_1=V'_1\oplus_1 V''_1$,
    \begin{align}
	&V=V_{1,2,3}=V^{(1)}\oplus_{1} V^{(2)}, \\
	&V^{(1)}=V'_1\otimes V_2\otimes V_3, \ V^{(2)}=V''_1\otimes V_2\otimes V_3,
	\label{}
    \end{align}
    the $\{1\}$ diagonal block subspaces $V'=V^{(1)}$, $V''=V^{(2)}$, and arbitrary vectors $v'\in V'$, $v''\in V''$, $v=v'+_{1}v''\in V$, we have
    \begin{align}
	&n_{1}(v'+_{1}v'')\ge\max\bigl\{d_{1}-d_{2}d_{3},n_{1}(v')+n_{1}(v'')\bigr\}, \label{n_1_direct_sum_1} \\
	&n_{2}(v'+_{1}v'')\ge\max\bigl\{0,n_{2}(v')+n_{2}(v'')-d_{2}\bigr\}, \label{n_2_direct_sum_1} \\
	&n_{3}(v'+_{1}v'')\ge\max\bigl\{0,n_{3}(v')+n_{3}(v'')-d_{3}\bigr\}, \label{n_3_direct_sum_1} \\
	&n_{1,2}(v'+_{1}v'')\ge\max\bigl\{d_{1}d_{2}-d_{3},n_{1,2}(v')+n_{1,2}(v'')\bigr\}, \label{n_1_2_direct_sum_1} \\
	&n_{1,3}(v'+_{1}v'')\ge\max\bigl\{d_{1}d_{3}-d_{2},n_{1,3}(v')+n_{1,3}(v'')\bigr\}, \label{n_1_3_direct_sum_1} \\
	&n_{2,3}(v'+_{1}v'')\ge\max\bigl\{0,n_{2,3}(v')+n_{2,3}(v'')-d_{2}d_{3}\bigr\}, \label{n_2_3_direct_sum_1} \\
	&n_{1,2,3}(v'+_{1}v'')\ge\max\bigl\{0,n_{1,2,3}(v')+n_{1,2,3}(v'')-d''_{1}(d'_{1}-n_{1}(v'))-d'_{1}(d''_{1}-n_{1}(v''))\bigr\}, \label{n_1_2_3_direct_sum_1}
    \end{align}
    where $d'_1=\dim{V'_1}$, $d''_1=\dim{V''_1}$.
    \label{direct_sum_lemma_1}
\end{lemmarep}
%------------------------------------------------------------------------------

\begin{toappendix}
    \label{appendix_direct_sum_lemma_1}
\end{toappendix}

See Appendix \ref{appendix_direct_sum_lemma_1} for the proof.

%------------------------------------------------------------------------------
\begin{appendixproof}
    See the first row of Figure~\ref{figure_direct_sum_1_2_3} for an illustration, from which it is clear that such a decomposition exists for an arbitrary $v\in V$.
    Equations \eqref{k_a}, \eqref{k_b_c} and \eqref{k_a_b_c} become
    \begin{align}
	&K_1(v'+_{1}v'')=\{ w'\in V'_1, \ w''\in V''_1 \colon C_1(v'\otimes w^{\prime *})+C_1(v''\otimes w^{\prime\prime *})=0 \}, \label{k_1_direct_sum_1} \\
	&K_2(v'+_{1}v'')=\{ w\in V_2 \colon C_2(v'\otimes w^*)=0, \ C_2(v''\otimes w^*)=0 \}, \label{k_2_direct_sum_1} \\
	&K_3(v'+_{1}v'')=\{ w\in V_3 \colon C_3(v'\otimes w^*)=0, \ C_3(v''\otimes w^*)=0 \}, \label{k_3_direct_sum_1} \\
	&K_{1,2}(v'+_{1}v'')=\{ w'\in V'_{1}\otimes V_{2}, \ w''\in V''_{1}\otimes V_{2} \colon C_{1}C_{2}(v'\otimes w^{\prime *})+C_{1}C_{2}(v''\otimes w^{\prime\prime *})=0 \}, \label{k_1_2_direct_sum_1} \\
	&K_{1,3}(v'+_{1}v'')=\{ w'\in V'_{1}\otimes V_{3}, \ w''\in V''_{1}\otimes V_{3} \colon C_{1}C_{3}(v'\otimes w^{\prime *})+C_{1}C_{3}(v''\otimes w^{\prime\prime *})=0 \}, \label{k_1_3_direct_sum_1} \\
	&K_{2,3}(v'+_{1}v'')=\{ w\in V_{2}\otimes V_{3} \colon C_{2}C_{3}(v'\otimes w^{*})=0, \ C_{2}C_{3}(v''\otimes w^{*})=0 \}, \label{k_2_3_direct_sum_1} \\
	&K_{1,2,3}(v'+_{1}v'')=\{ w'\in V', \ w''\in V'' \colon C_1 C_2(v'\otimes w^{\prime *})+C_1 C_2(v''\otimes w^{\prime\prime *})=0, \nn \\
	    &\quad C_1 C_3(v'\otimes w^{\prime *}) +C_1 C_3(v''\otimes w^{\prime\prime *})=0, \ C_2 C_3(v'\otimes w^{\prime *})=0, \ C_2 C_3(v'\otimes w^{\prime\prime *})=0, \nn \\
	&\quad C_2 C_3(v''\otimes w^{\prime *})=0, \ C_2 C_3(v''\otimes w^{\prime\prime *})=0 \}
	\label{k_1_2_3_direct_sum_1}
    \end{align}
    and the lemma follows directly by using \eqref{inequalities_for_systems_of_linear_equations}. 
    For example, for \eqref{n_1_2_3_direct_sum_1-apx} we have
    \begin{align}
	&n_{1,2,3}(v'+_{1}v'')=d_{1} d_{2} d_{3}-N\bigl[w'\in V',w''\in V''\colon \nn \\
	    &\quad C_{1}C_{2}(v'\otimes w^{\prime *})+C_{1}C_{2}(v''\otimes w^{\prime\prime *})=0,C_{1}C_{3}(v'\otimes w^{\prime *})+C_{1}C_{3}(v''\otimes w^{\prime\prime *})=0, \nn \\
	&\quad C_{2}C_{3}(v'\otimes w^{\prime *})=0,C_{2}C_{3}(v'\otimes w^{\prime\prime *})=0,C_{2}C_{3}(v''\otimes w^{\prime *})=0,C_{2}C_{3}(v''\otimes w^{\prime\prime *})=0\bigr] \nn \\
	&\quad \ge d_{1}d_{2}d_{3} \nn \\
	&\quad -\min\Bigl\{d_{1}d_{2}d_{3},N\bigl[ w'\in V' \colon C_{1} C_{2}(v'\otimes w^{\prime *})=0, \ C_{1} C_{3}(v'\otimes w^{\prime *})=0, \ C_{2} C_{3}(v'\otimes w^{\prime *})=0 \bigr] \nn \\
	    &\quad+N\bigl[ w''\in V'' \colon C_{1} C_{2}(v''\otimes w^{\prime\prime *})=0, \ C_{1} C_{3}(v''\otimes w^{\prime\prime *})=0, \ C_{2} C_{3}(v''\otimes w^{\prime\prime *})=0 \bigr] \nn \\
	&\quad +N\bigl[ w''\in V'' \colon C_{2} C_{3}(v'\otimes w^{\prime\prime *})=0 \bigr] -N\bigl[ w'\in V' \colon C_{2} C_{3}(v''\otimes w^{\prime *})=0 \bigr]\Bigr\} \nn \\
	&\quad =d_{1}d_{2}d_{3}-\min\Bigl\{d_{1}d_{2}d_{3},(d'_{1} d_{2} d_{3} -n_{1,2,3}(v')) +(d''_{1} d_{2} d_{3} -n_{1,2,3}(v'')) \nn \\
	&\quad +N\bigl[ w''\in V'' \colon C_{2} C_{3}(v'\otimes w^{\prime\prime *})=0 \bigr] +N\bigl[ w'\in V' \colon C_{2} C_{3}(v''\otimes w^{\prime *})=0 \bigr]\Bigr\}.
	\label{}
    \end{align}
    We now use 
    \begin{align}
	&N\bigl[ w''\in V'' \colon C_{2} C_{3}(v'\otimes w^{\prime\prime *})=0 \bigr]=\min\bigl\{d'_{1}d''_{1},d''_{1}(d_{2}d_{3}-n_{2,3}(v'))\bigr\} \nn \\
	&\quad =\min\bigl\{d'_{1}d''_{1},d''_{1}(d'_{1}-n_{1}(v'))\bigr\} =d''_{1}(d'_{1}-n_{1}(v')), \\
	&N\bigl[ w'\in V' \colon C_{2} C_{3}(v''\otimes w^{\prime *})=0 \bigr]=\min\bigl\{d'_{1}d''_{1},d'_{1}(d_{2}d_{3}-n_{2,3}(v''))\bigr\} \nn \\
	&\quad =\min\bigl\{d'_{1}d''_{1},d'_{1}(d''_{1}-n_{1}(v''))\bigr\}=d'_{1}(d''_{1}-n_{1}(v'')), \label{}
    \end{align}
    which follow from the definitions of $n_{2,3}(v')$ and $n_{2,3}(v'')$ together with the requirement that the number of constraints cannot exceed the number of equations specifying the constraints, and finally the relation \eqref{n_a_n_b_c}.
    As a result,
    \begin{align}
	&n_{1,2,3}(v'+_{1}v'')\ge\max\bigl\{0,n_{1,2,3}(v')+n_{1,2,3}(v'')-d''_{1}(d'_{1}-n_{1}(v'))-d'_{1}(d''_{1}-n_{1}(v''))\bigr\}.
	\label{}
    \end{align}
\end{appendixproof}
%------------------------------------------------------------------------------

Instead of decomposing $V_1$ into direct sum components, as we did above, we can also choose to decompose either $V_2$ or $V_3$.
The corresponding results easily follow from the above equations by permutation of indices.

%------------------------------------------------------------------------------
\subsubsection{Two decompositions} \label{section_direct_sum_1_2}
%------------------------------------------------------------------------------

%------------------------------------------------------------------------------
\begin{lemmarep}
    For a direct sum decomposition generated by $V_a=V'_a\oplus_a V''_a$, $1\le a\le 2$,
    \begin{align}
	&V=V_{1,2,3}=V^{(1)}\oplus_{1,2} \dotsb \oplus_{1,2} V^{(4)}, \\
	&V^{(1)}=V'_1\otimes V'_2\otimes V_3, \ V^{(2)}=V'_1\otimes V''_2\otimes V_3, \ V^{(3)}=V''_1\otimes V'_2\otimes V_3, \ V^{(4)}=V''_1\otimes V''_2\otimes V_3,
	\label{}
    \end{align}
    the $\{1,2\}$ diagonal block subspaces $V'=V^{(1)}$, $V''=V^{(4)}$, and arbitrary vectors $v'\in V'$, $v''\in V''$, $v=v'+_{1,2}v''\in V$, we have
    \begin{align}
	&n_{1}(v'+_{1,2}v'')=n_{1}(v')+n_{1}(v''), \label{n_1_direct_sum_1_2} \\
	&n_{2}(v'+_{1,2}v'')=n_{2}(v')+n_{2}(v''), \label{n_2_direct_sum_1_2} \\
	&n_{3}(v'+_{1,2}v'')\ge\max\bigl\{0,n_{3}(v')+n_{3}(v'')-d_{3}\bigr\}, \label{n_3_direct_sum_1_2} \\
	&n_{1,2}(v'+_{1,2}v'')\ge\max\bigl\{d_{1}d_{2}-d_{3},n_{1,2}(v')+n_{1,2}(v'')+d'_{1}d''_{2}+d''_{1}d'_{2}\bigr\}, \label{n_1_2_direct_sum_1_2} \\
	&n_{1,3}(v'+_{1,2}v'')=n_{1,3}(v')+n_{1,3}(v''), \label{n_1_3_direct_sum_1_2} \\
	&n_{2,3}(v'+_{1,2}v'')=n_{2,3}(v')+n_{2,3}(v''), \label{n_2_3_direct_sum_1_2} \\
	&n_{1,2,3}(v'+_{1,2}v'')\ge\max\bigl\{0,n_{1,2,3}(v')+n_{1,2,3}(v'')-(d'_{1} d''_{2}+d''_{1} d'_{2})d_{3} \nn \\
	&\quad +d''_{1}n_{2,3}(v')+d''_{2}n_{1,3}(v')+d'_{1}n_{2,3}(v'')+d'_{2}n_{1,3}(v'')\bigr\}, \label{n_1_2_3_direct_sum_1_2}
    \end{align}
    where $d'_{a}=\dim{V'_{a}}$, $d''_{a}=\dim{V''_{a}}$, $1\le a\le 2$.
    \label{direct_sum_lemma_1_2}
\end{lemmarep}
%------------------------------------------------------------------------------

\begin{toappendix}
    \label{appendix_direct_sum_lemma_1_2}
\end{toappendix}

See Appendix \ref{appendix_direct_sum_lemma_1_2} for the proof.

%------------------------------------------------------------------------------
\begin{appendixproof}
    See the second row of Figure~\ref{figure_direct_sum_1_2_3} for an illustration, from which it is clear that such a decomposition exists only for certain vectors $v\in V$.
    Equations \eqref{k_a}, \eqref{k_b_c} and \eqref{k_a_b_c} become
    \begin{align}
	&K_{1}(v'+_{1,2}v'')=K_{1}(v')\cup K_{1}(v''), \label{k_1_direct_sum_1_2} \\
	&K_{2}(v'+_{1,2}v'')=K_{2}(v')\cup K_{2}(v''), \label{k_2_direct_sum_1_2} \\
	&K_{3}(v'+_{1,2}v'')=\{ w\in V_{3} \colon C_{3}(v'\otimes w^*)=0, \ C_{3}(v''\otimes w^*)=0 \}, \label{k_3_direct_sum_1_2} \\
	&K_{1,2}(v'+_{1,2}v'')=\{ w'\in V'_{1}\otimes V'_{2}, \ w''\in V''_{1}\otimes V''_{2} \colon C_{1} C_{2}(v'\otimes w^{\prime *})+C_{1} C_{2}(v''\otimes w^{\prime\prime *})=0 \}, \label{k_1_2_direct_sum_1_2} \\
	&K_{1,3}(v'+_{1,2}v'')=K_{1,3}(v')\cup K_{1,3}(v''), \label{k_1_3_direct_sum_1_2} \\
	&K_{2,3}(v'+_{1,2}v'')=K_{2,3}(v')\cup K_{2,3}(v''), \label{k_2_3_direct_sum_1_2} \\
	&K_{1,2,3}(v'+_{1,2}v'')=\{ w_{1}\in V^{(1)}, \dotsc, w_4\in V^{(4)} \colon C_{1} C_{2}(v'\otimes w_{1}^*)+C_{1} C_{2}(v''\otimes w_4^*)=0, \nn \\
	    &\quad C_{1} C_{3}(v'\otimes w_{1}^*)=0,\ C_{1} C_{3}(v'\otimes w_{2}^*)=0, \ C_{1} C_{3}(v''\otimes w_{3}^*)=0, \ C_{1} C_{3}(v''\otimes w_4^*)=0, \nn \\
	&\quad C_{2} C_{3}(v'\otimes w_{1}^*)=0, \ C_{2} C_{3}(v'\otimes w_{3}^*)=0, \ C_{2} C_{3}(v''\otimes w_{2}^*)=0, \ C_{2} C_{3}(v''\otimes w_4^*)=0 \} \label{k_1_2_3_direct_sum_1_2}  
    \end{align}
    and the lemma follows directly by using \eqref{inequalities_for_systems_of_linear_equations}. 
    For example, for \eqref{n_1_2_3_direct_sum_1_2-apx} we have
    \begin{align}
	&n_{1,2,3}(v'+_{1,2}v'')=d_{1}d_{2}d_{3}-N\bigl[ w_{1}\in V^{(1)}, \dotsc ,w_4\in V^{(4)} \colon \nn \\
	    &\quad C_{1}C_{2}(v'\otimes w_{1}^{*})+C_{1}C_{2}(v''\otimes w_4^{*})=0, \nn \\
	    &\quad C_{1}C_{3}(v'\otimes w_{1}^{*})=0,\ C_{1}C_{3}(v'\otimes w_{2}^{*})=0, \ C_{1}C_{3}(v''\otimes w_{3}^{*})=0, \ C_{1}C_{3}(v''\otimes w_4^{*})=0, \nn \\
	&\quad C_{2}C_{3}(v'\otimes w_{1}^{*})=0, \ C_{2}C_{3}(v'\otimes w_{3}^{*})=0, \ C_{2}C_{3}(v''\otimes w_{2}^{*})=0, \ C_{2}C_{3}(v''\otimes w_4^{*})=0\bigr] \nn \\
	&\quad \ge d_{1}d_{2}d_{3} \nn \\
	&\quad -\min\Bigl\{d_{1}d_{2}d_{3},N\bigl[ w_{1}\in V^{(1)}\colon C_{1}C_{2}(v'\otimes w^{*}_{1})=0, \ C_{1}C_{3}(v'\otimes w^{*}_{1})=0, \ C_{2}C_{3}(v'\otimes w^{*}_{1})=0 \bigr] \nn \\
	    &\quad +N\bigl[ w_4\in V^{(4)}\colon C_{1}C_{2}(v''\otimes w^{*}_4)=0, \ C_{1}C_{3}(v''\otimes w^{*}_4)=0, \ C_{2}C_{3}(v''\otimes w^{*}_4)=0 \bigr] \nn \\
	    &\quad +N\bigl[ w_{2}\in V^{(2)}\colon C_{1}C_{3}(v'\otimes w^{*}_{2})=0, \ C_{2}C_{3}(v''\otimes w^{*}_{2})=0 \bigr] \nn \\
	&\quad +N\bigl[ w_{3}\in V^{(3)}\colon C_{1}C_{3}(v''\otimes w^{*}_{3})=0, \ C_{2}C_{3}(v'\otimes w^{*}_{3})=0 \bigr]\Bigr\} \nn \\
	&\quad =d_{1}d_{2}d_{3}-\min\Bigl\{d_{1}d_{2}d_{3},(d'_{1}d'_{2}d_{3}-n_{1,2,3}(v'))+(d''_{1}d''_{2}d_{3}-n_{1,2,3}(v'')) \nn \\
	    &\quad +N\bigl[ w_{2}\in V^{(2)}\colon C_{1}C_{3}(v'\otimes w^{*}_{2})=0, \ C_{2}C_{3}(v''\otimes w^{*}_{2})=0 \bigr] \nn \\
	&\quad +N\bigl[ w_{3}\in V^{(3)}\colon C_{1}C_{3}(v''\otimes w^{*}_{3})=0, \ C_{2}C_{3}(v'\otimes w^{*}_{3})=0 \bigr]\Bigr\} \label{}
    \end{align}
    We now use
    \begin{align}
	&N\bigl[ w_{2}\in V^{(2)}\colon C_{1}C_{3}(v'\otimes w^{*}_{2})=0, \ C_{2}C_{3}(v''\otimes w^{*}_{2})=0 \bigr] \nn \\
	&\quad \le\min\bigl\{d'_{1}d''_{1}+d'_{2}d''_{2},d''_{2}(d'_{1}d_{3}-n_{1,3}(v'))+d'_{1}(d''_{2}d_{3}-n_{2,3}(v''))\bigr\} \nn \\
	&\quad =d''_{2}(d'_{1}d_{3}-n_{1,3}(v'))+d'_{1}(d''_{2}d_{3}-n_{2,3}(v'')), \label{direct_sum_lemma_1_2_N_w_2}\\
	&N\bigl[ w_{3}\in V^{(3)}\colon C_{1}C_{3}(v''\otimes w^{*}_{3})=0, \ C_{2}C_{3}(v'\otimes w^{*}_{3})=0 \bigr] \nn \\
	&\quad \le\min\bigl\{d'_{1}d''_{1}+d'_{2}d''_{2},d''_{1}(d'_{2}d_{3}-n_{2,3}(v'))+d'_{2}(d''_{1}d_{3}-n_{1,3}(v''))\bigr\} \nn \\
	&\quad =d''_{1}(d'_{2}d_{3}-n_{2,3}(v'))+d'_{2}(d''_{1}d_{3}-n_{1,3}(v'')), \label{direct_sum_lemma_1_2_N_w_4}
    \end{align}
    which follow from the definitions of $n_{1,3}(v')$, $n_{2,3}(v'')$, $n_{2,3}(v')$ and $n_{1,3}(v'')$  together with the requirement that the number of constraints cannot exceed the number of equations specifying the constraints.
    The equalities in \eqref{direct_sum_lemma_1_2_N_w_2} and \eqref{direct_sum_lemma_1_2_N_w_4} follow from \eqref{n_a_n_b_c}, which gives
    \begin{align}
	&d'_{1}d_{3}-n_{1,3}(v')=d'_{2}-n_{2}(v')\le d'_{2}, \\
	&d''_{2}d_{3}-n_{2,3}(v'')=d''_{1}-n_{1}(v'')\le d''_{1}, \\
	&d'_{2}d_{3}-n_{2,3}(v')=d'_{1}-n_{1}(v')\le d'_{1}, \\
	&d''_{1}d_{3}-n_{1,3}(v'')=d''_{2}-n_{2}(v'')\le d''_{2}.
	\label{}
    \end{align}
    As a result,
    \begin{align}
	&n_{1,2,3}(v'+_{1,2}v'')\ge\max\bigl\{0,n_{1,2,3}(v')+n_{1,2,3}(v'')-(d'_{1} d''_{2}+d''_{1} d'_{2})d_{3} \nn \\
	&\quad +d''_{1}n_{2,3}(v')+d''_{2}n_{1,3}(v')+d'_{1}n_{2,3}(v'')+d'_{2}n_{1,3}(v'')\bigr\}.  \label{}
    \end{align}
\end{appendixproof}
%------------------------------------------------------------------------------

Since the direct sum decomposition of two spaces imply such decompositions for each of the two involved spaces, it follows that \eqref{n_1_direct_sum_1_2}, \eqref{n_2_direct_sum_1_2}, \eqref{n_3_direct_sum_1_2}, \eqref{n_1_2_direct_sum_1_2}, \eqref{n_1_3_direct_sum_1_2}, \eqref{n_2_3_direct_sum_1_2}, \eqref{n_1_2_3_direct_sum_1_2} imply \eqref{n_1_direct_sum_1}, \eqref{n_2_direct_sum_1}, \eqref{n_3_direct_sum_1}, \eqref{n_1_2_direct_sum_1}, \eqref{n_1_3_direct_sum_1}, \eqref{n_2_3_direct_sum_1}, \eqref{n_1_2_3_direct_sum_1} and their counterparts for the decomposition of $V_2$ instead of $V_1$ in Section~\ref{section_direct_sum_1}. 

Instead of decomposing $V_1$, $V_2$ into direct sum components, as we did above, we can also choose to decompose either $V_1$, $V_3$ or $V_2$, $V_3$.
The corresponding results easily follow from the above equations by permutation of indices.

%------------------------------------------------------------------------------
\subsubsection{Three decompositions} \label{section_direct_sum_1_2_3}
%------------------------------------------------------------------------------

%------------------------------------------------------------------------------
\begin{lemmarep}
    For a direct sum decomposition generated by $V_a=V'_a\oplus_a V''_a$, $1\le a\le 3$,
    \begin{align}
	&V=V_{1,2,3}=V^{(1)}\oplus_{1,2,3} \dotsb \oplus_{1,2,3} V^{(8)}, \\
	&V^{(1)}=V'_1\otimes V'_2\otimes V'_3, \ V^{(2)}=V'_1\otimes V'_2\otimes V''_3, \ V^{(3)}=V'_1\otimes V''_2\otimes V'_3, \ V^{(4)}=V'_1\otimes V''_2\otimes V''_3 \nn \\
	&V^{(5)}=V''_1\otimes V'_2\otimes V'_3, \ V^{(6)}=V''_1\otimes V'_2\otimes V''_3, \ V^{(7)}=V''_1\otimes V''_2\otimes V'_3, \ V^{(8)}=V''_1\otimes V''_2\otimes V''_3,
	\label{}
    \end{align}
    the $\{1,2,3\}$ diagonal block subspaces $V'=V^{(1)}$, $V''=V^{(8)}$, and arbitrary vectors $v'\in V'$, $v''\in V''$, $v=v'+_{1,2,3}v''\in V$, we have
    \begin{align}
	&n_{a}(v'+_{1,2,3}v'')=n_{a}(v')+n_{a}(v''), \ 1\le a\le 3, \label{n_a_direct_sum_1_2_3} \\
	&n_{b,c}(v'+_{1,2,3}v'')=n_{b,c}(v')+n_{b,c}(v'')+d'_{b}d''_{c}+d''_{b}d'_{c}, \ 1\le b,c\le 3, \ b\not=c, \label{n_b_c_direct_sum_1_2_3} \\
	&n_{1,2,3}(v'+_{1,2,3}v'')=n_{1,2,3}(v')+n_{1,2,3}(v'')+d''_{1}n_{2,3}(v')+d''_{2}n_{1,3}(v')+d''_{3}n_{1,2}(v') \nn \\
	&\quad +d'_{1}n_{2,3}(v'')+d'_{2}n_{1,3}(v'')+d'_{3}n_{1,2}(v''), \label{n_1_2_3_direct_sum_1_2_3}
    \end{align}
    where $d'_a=\dim{V'_a}$, $d''_a=\dim{V''_a}$, $1\le a\le 3$.
    \label{direct_sum_lemma_1_2_3}
\end{lemmarep}
%------------------------------------------------------------------------------

\begin{toappendix}
    \label{appendix_direct_sum_lemma_1_2_3}
\end{toappendix}

See Appendix \ref{appendix_direct_sum_lemma_1_2_3} for the proof.

%------------------------------------------------------------------------------
\begin{appendixproof}
    See the third row of Figure~\ref{figure_direct_sum_1_2_3} for an illustration, from which it is clear that such a decomposition exists only for certain vectors $v\in V$.
    Equations \eqref{k_a}, \eqref{k_b_c} and \eqref{k_a_b_c} become
    \begin{align}
	&K_{a}(v'+_{1,2,3}v'')=K_{a}(v')\cup K_{a}(v''), \ 1\le a\le 3, \label{k_a_direct_sum_1_2_3} \\
	&K_{b,c}(v'+_{1,2,3}v'')=K_{b,c}(v')\cup K_{b,c}(v''), \ 1\le b,c\le 3, \ b\not=c, \label{k_b_c_direct_sum_1_2_3} \\
	&K_{1,2,3}(v'+_{1,2,3}v'')=K_{1,2,3}(v')\cup K_{1,2,3}(v'') \nn \\
	&\quad \cup \{w_2\in V^{(2)} \colon C_1 C_2(v'\otimes w^*_2)=0\} \cup \{w_3\in V^{(3)} \colon C_1 C_3(v'\otimes w^*_3)=0\} \nn \\
	&\quad \cup \{w_4\in V^{(4)} \colon C_2 C_3(v''\otimes w^*_4)=0\} \cup \{w_5\in V^{(5)} \colon C_2 C_3(v'\otimes w^*_{5})=0\}, \nn \\ 
	&\quad \cup \{w_6\in V^{(6)} \colon C_1 C_3(v''\otimes w^*_6)=0\} \cup \{w_7\in V^{(7)} \colon C_1 C_2(v''\otimes w^*_7)=0\}
	\label{k_1_2_3_direct_sum_1_2_3}
    \end{align}
    and the lemma follows directly. 
    For example, \eqref{n_1_2_3_direct_sum_1_2_3-apx} follows from
    \begin{align}
	&n_{1,2,3}(v'+_{1,2,3}v'')=n_{1,2,3}(v')+n_{1,2,3}(v'') \nn \\
	&\quad +d'_{1}d'_{2}d''_{3}-N\bigl[ w_{2}\in V^{(2)}\colon C_{1}C_{2}(v'\otimes w_{2}^{*})=0 \bigr] \nn \\
	&\quad +d'_{1}d''_{2}d'_{3}-N\bigl[ w_{3}\in V^{(3)}\colon C_{1}C_{3}(v'\otimes w_{3}^{*})=0 \bigr] \nn \\
	&\quad +d'_{1}d''_{2}d''_{3}-N\bigl[ w_{4}\in V^{(4)}\colon C_{2}C_{3}(v''\otimes w_{4}^{*})=0 \bigr] \nn \\
	&\quad +d''_{1}d'_{2}d'_{3}-N\bigl[ w_{5}\in V^{(5)}\colon C_{2}C_{3}(v'\otimes w_{5}^{*})=0 \bigr] \nn \\
	&\quad +d''_{1}d'_{2}d''_{3}-N\bigl[ w_{6}\in V^{(6)}\colon C_{1}C_{3}(v''\otimes w_{6}^{*})=0 \bigr] \nn \\
	&\quad +d''_{1}d''_{2}d'_{3}-N\bigl[ w_{7}\in V^{(7)}\colon C_{1}C_{2}(v''\otimes w_{7}^{*})=0 \bigr] \nn \\
	&\quad =n_{1,2,3}(v')+n_{1,2,3}(v'')+d''_{3}n_{1,2}(v')+d''_{2}n_{1,3}(v')+d'_{1}n_{2,3}(v'') \nn \\
	&\quad +d''_{1}n_{2,3}(v')+d'_{2}n_{1,3}(v'')+d'_{3}n_{1,2}(v'').
	\label{}
    \end{align}
\end{appendixproof}
%------------------------------------------------------------------------------

We note that the direct sum operation for vector spaces is commutative and associative as follows from the same properties of the addition of vectors.
The commutativity is also explicit in \eqref{n_a_direct_sum_1_2_3}, \eqref{n_b_c_direct_sum_1_2_3}, \eqref{n_1_2_3_direct_sum_1_2_3}, and using these equations it is easy to prove the associativity.
For example, the associativity for $n_{1,2,3}$ follows from
\begin{align}
    &n_{1,2,3}((v'+_{1,2,3}v'')+_{1,2,3}v''')=n_{1,2,3}(v'+_{1,2,3}v'')+n_{1,2,3}(v''') \nn \\
    &\quad +d'''_{1}n_{2,3}(v'+_{1,2,3}v'')+d'''_{2}n_{1,3}(v'+_{1,2,3}v'')+d'''_{3}n_{1,2}(v'+_{1,2,3}v'') \nn \\
    &\quad +(d'_{1}+d''_{1})n_{2,3}(v''')+(d'_{2}+d''_{2})n_{1,3}(v''')+(d'_{3}+d''_{3})n_{1,2}(v''') \nn \\
    &\quad =n_{1,2,3}(v')+n_{1,2,3}(v'')+d''_{1}n_{2,3}(v')+d''_{2}n_{1,3}(v')+d''_{3}n_{1,2}(v'), \nn \\
    &\quad +d'_{1}n_{2,3}(v'')+d'_{2}n_{1,3}(v'')+d'_{3}n_{1,2}(v'')+n_{1,2,3}(v''') \\
%    &n_{1,2,3}(v'+_{1,2,3}(v''+_{1,2,3}v'''))=n_{1,2,3}(v')+n_{1,2,3}(v''+_{1,2,3}v''') \nn \\
    &\quad +d'''_{1}\bigl(n_{2,3}(v')+n_{2,3}(v'')+d'_{2}d''_{3}+d''_{2}d'_{3}\bigr) \nn \\
    &\quad +d'''_{2}\bigl(n_{1,3}(v')+n_{1,3}(v'')+d'_{1}d''_{3}+d''_{1}d'_{3}\bigr) \nn \\
    &\quad +d'''_{3}\bigl(n_{1,2}(v')+n_{1,2}(v'')+d'_{1}d''_{2}+d''_{1}d'_{2}\bigr) \nn \\
    &\quad +(d'_{1}+d''_{1})n_{2,3}(v''')+(d'_{2}+d''_{2})n_{1,3}(v''')+(d'_{3}+d''_{3})n_{1,2}(v'''), \\
    &n_{1,2,3}(v'+_{1,2,3}(v''+_{1,2,3}v'''))=n_{1,2,3}(v')+n_{1,2,3}(v''+_{1,2,3}v''') \nn \\
    &\quad +(d''_{1}+d'''_{1})n_{2,3}(v')+(d''_{2}+d'''_{2})n_{1,3}(v')+(d''_{3}+d'''_{3})n_{1,2}(v') \nn \\
    &\quad +d'_{1}n_{2,3}(v''+_{1,2,3}v''')+d'_{2}n_{1,3}(v''+_{1,2,3}v''')+d'_{3}n_{1,2}(v''+_{1,2,3}v''') \nn \\
    &\quad =n_{1,2,3}(v')+n_{1,2,3}(v'')+n_{1,2,3}(v''')+d'''_{1}n_{2,3}(v'')+d'''_{2}n_{1,3}(v'')+d'''_{3}n_{1,2}(v''), \nn \\
    &\quad +d''_{1}n_{2,3}(v''')+d''_{2}n_{1,3}(v''')+d''_{3}n_{1,2}(v''') \\
    &\quad +(d''_{1}+d'''_{1})n_{2,3}(v')+(d''_{2}+d'''_{2})n_{1,3}(v')+(d''_{3}+d'''_{3})n_{1,2}(v'), \\
    &\quad +d'_{1}\bigl(n_{2,3}(v'')+n_{2,3}(v''')+d''_{2}d'''_{3}+d'''_{2}d''_{3}\bigr) \nn \\
    &\quad +d'_{2}\bigl(n_{1,3}(v'')+n_{1,3}(v''')+d''_{1}d'''_{3}+d'''_{1}d''_{3}\bigr) \nn \\
    &\quad +d'_{3}\bigl(n_{1,2}(v'')+n_{1,2}(v''')+d''_{1}d'''_{2}+d'''_{1}d''_{2}\bigr)
    \label{}
\end{align}
for any vectors $v'\in V'$, $v''\in V''$, $v'''\in V'''$, where
\begin{align}
    &V'=V'_1\otimes V'_2\otimes V'_3, \ V''=V''_1\otimes V''_2\otimes V''_3, \ V'''=V'''_1\otimes V'''_2\otimes V'''_3, \\
    &V_a=V'_a\oplus V''_a\oplus V'''_a, \ 1\le a\le 3.
    \label{}
\end{align}

Now Lemma \ref{direct_sum_lemma_1_2_3} leads to the following decomposition theorem.

%------------------------------------------------------------------------------
\begin{theoremrep}
    For the direct sum decompositions $V_a=(\oplus_a)_{q=1}^p V_a^{(q)}$, $1\le a\le 3$, $p\ge 1$, let $V^{(q)}=V_{1}^{(q)}\otimes V_{2}^{(q)}\otimes V_{3}^{(q)}$, $1\le q\le p$ be the $\{1,2,3\}$ diagonal blocks.
    The algebraic invariants of $v=(+_{1,2,3})_{q=1}^p v^{(q)}=v^{(1)}+_{1,2,3}\dotsb+_{1,2,3}v^{(p)}$ can be found in terms of the corresponding invariants of its components $v^{(q)}\in V^{(q)}$, $1\le q\le p$ as follows:
    \begin{align}
	&n_{a}\bigl((+_{1,2,3})_{q=1}^{p}v^{(q)}\bigr)=\sum_{q=1}^{p} n_{a}(v^{(q)}), \ 1\le a\le 3, \label{theorem_n_a} \\
	&n_{b,c}\bigl((+_{1,2,3})_{q=1}^{p}v^{(q)}\bigr)=\sum_{q=1}^{p}n_{b,c}(v^{(q)})+\sum_{
	    \begin{subarray}{c}
		1\le q_{1},q_{2}\le p \\
		q_{1}\not=q_{2}
	    \end{subarray}
	}d_{b}^{(q_{1})}d_{c}^{(q_{2})}, \ 1\le b<c\le 3, \label{theorem_n_b_c} \\
	&n_{1,2,3}\bigl((+_{1,2,3})_{q=1}^{p} v^{(q)}\bigr)=\sum_{q=1}^{p} n_{1,2,3}(v^{(q)}) \nn \\
	&\quad +\Bigl(\sum_{q=1}^{p}d_{1}^{(q)}\Bigr)\Bigl(\sum_{q=1}^{p}n_{2,3}(v^{(q)})\Bigr)+\Bigl(\sum_{q=1}^{p}d_{2}^{(q)}\Bigr)\Bigl(\sum_{q=1}^{p}n_{1,3}(v^{(q)})\Bigr)+\Bigl(\sum_{q=1}^{p}d_{3}^{(q)}\Bigr)\Bigl(\sum_{q=1}^{p}n_{1,2}(v^{(q)})\Bigr) \nn \\
	&\quad -\sum_{q=1}^{p}d_{1}^{(q)}n_{2,3}(v^{(q)})-\sum_{q=1}^{p}d_{2}^{(q)}n_{1,3}(v^{(q)})-\sum_{q=1}^{p}d_{3}^{(q)}n_{1,2}(v^{(q)}) \nn \\
	&\quad +\sum_{
	    \begin{subarray}{c}
		1\le q_{1},q_{2},q_{3}\le p, \\
		q_{1}\not=q_{2},\, q_{2}\not=q_{3},\, q_{3}\not=q_{1}
	    \end{subarray}
	}d_{1}^{(q_{1})}d_{2}^{(q_{2})}d_{3}^{(q_{3})}, \label{theorem_n_1_2_3}
    \end{align}
    where $d_a^{(q)}=\dim{V_a^{(q)}}$.
    \label{direct_sum_theorem}
\end{theoremrep}
%------------------------------------------------------------------------------

\begin{toappendix}
    \label{appendix_direct_sum_theorem}
\end{toappendix}

See Appendix \ref{appendix_direct_sum_theorem} for the proof.

%------------------------------------------------------------------------------
\begin{appendixproof}
    We proceed by induction.
    The statement of the theorem is trivially true for $p=1$ and \eqref{n_a_direct_sum_1_2_3}, \eqref{n_b_c_direct_sum_1_2_3} and \eqref{n_1_2_3_direct_sum_1_2_3} prove it for $p=2$.

    For the inductive step, we assume the statement holds for an arbitrary $p$ and use \eqref{n_a_direct_sum_1_2_3}, \eqref{n_b_c_direct_sum_1_2_3} and \eqref{n_1_2_3_direct_sum_1_2_3} to prove it for $p+1$.
    For \eqref{theorem_n_a-apx}, the inductive step follows from
    \begin{align}
	&n_{a}\bigl((+_{1,2,3})_{q=1}^{p+1}v^{(q)}\bigr)=n_{a}\bigl((+_{1,2,3})_{q=1}^{p}v^{(q)}+_{1,2,3}v^{(p+1)}\bigr)=\sum_{q=1}^{p}n_{a}(v^{(q)})+n_{a}(v^{(p+1)}) \nn \\
	&\quad =\sum_{q=1}^{p+1}n_{a}(v^{(q)}), \ 1\le a\le 3.
    \end{align}
    For \eqref{theorem_n_b_c-apx}, the inductive step follows from
    \begin{align}
	&n_{b,c}\bigl((+_{1,2,3})_{q=1}^{p+1}v^{(q)}\bigr)=n_{b,c}\bigl((+_{1,2,3})_{q=1}^{p}v^{(q)}+_{1,2,3}v^{(p+1)}\bigr) \nn \\
	&\quad =n_{b,c}\bigl((+_{1,2,3})_{q=1}^{p}v^{(q)}\bigl)+n_{b,c}(v^{(p+1)})+\Bigl(\sum_{q=1}^{p}d_{b}^{(q)}\Bigr)d_{c}^{(p+1)}+d_{b}^{(p+1)}\Bigl(\sum_{q=1}^{p}d_{c}^{(q)}\Bigr) \nn \\
	&\quad =\sum_{q=1}^{p}n_{b,c}(v^{(q)})+\sum_{
	    \begin{subarray}{c}
		1\le q_{1},q_{2}\le p \\
		q_{1}\not=q_{2}
	    \end{subarray}
	}d_{b}^{(q_{1})}d_{c}^{(q_{2})}+n_{b,c}(v^{(p+1)})+\Bigl(\sum_{q=1}^{p}d_{b}^{(q)}\Bigr)d_{c}^{(p+1)}+d_{b}^{(p+1)}\Bigl(\sum_{q=1}^{p}d_{c}^{(q)}\Bigr), \nn \\
	&\quad =\sum_{q=1}^{p+1}n_{b,c}(v^{(q)})+\sum_{
	    \begin{subarray}{c}
		1\le q_{1},q_{2}\le p+1 \\
		q_{1}\not=q_{2}
	    \end{subarray}
	}d_{b}^{(q_{1})}d_{c}^{(q_{2})}, \ 1\le b<c\le 3. \label{}
    \end{align}
    Finally, for \eqref{theorem_n_1_2_3-apx}, the inductive step follows from
    \begin{align}
	&n_{1,2,3}\bigl((+_{1,2,3})_{q=1}^{p+1}v^{(q)}\bigr) \nn \\
	&\quad =n_{1,2,3}\bigl((+_{1,2,3})_{q=1}^{p}v^{(q)}+_{1,2,3}v^{(p+1)}\bigr) \nn \\
	&\quad =n_{1,2,3}\bigl((+_{1,2,3})_{q=1}^{p}v^{(q)}\bigr)+d_{1}^{(p+1)}n_{2,3}\bigl((+_{1,2,3})_{q=1}^{p}v^{(q)}\bigr)+d_{2}^{(p+1)}n_{1,3}\bigl((+_{1,2,3})_{q=1}^{p}v^{(q)}\bigr) \nn \\
	&\quad +d_{3}^{(p+1)}n_{1,2}\bigl((+_{1,2,3})_{q=1}^{p}v^{(q)}\bigr)+\Bigl(\sum_{q=1}^{p}d_{1}^{(q)}\Bigr)n_{2,3}(v^{(p+1)})+\Bigl(\sum_{q=1}^{p}d_{2}^{(q)}\Bigr)n_{1,3}(v^{(p+1)}) \nn \\
	&\quad +\Bigl(\sum_{q=1}^{p}d_{3}^{(q)}\Bigr)n_{1,2}(v^{(p+1)})+n_{1,2,3}(v^{(p+1)}) \nn \\
	&\quad =\sum_{q=1}^{p} n_{1,2,3}(v^{(q)}) \nn \\
	&\quad +\Bigl(\sum_{q=1}^{p}d_{1}^{(q)}\Bigr)\Bigl(\sum_{q=1}^{p}n_{2,3}(v^{(q)})\Bigr)+\Bigl(\sum_{q=1}^{p}d_{2}^{(q)}\Bigr)\Bigl(\sum_{q=1}^{p}n_{1,3}(v^{(q)})\Bigr)+\Bigl(\sum_{q=1}^{p}d_{3}^{(q)}\Bigr)\Bigl(\sum_{q=1}^{p}n_{1,2}(v^{(q)})\Bigr) \nn \\
	&\quad -\sum_{q=1}^{p}d_{1}^{(q)}n_{2,3}(v^{(q)})-\sum_{q=1}^{p}d_{2}^{(q)}n_{1,3}(v^{(q)})-\sum_{q=1}^{p}d_{3}^{(q)}n_{1,2}(v^{(q)}) \nn \\
	&\quad +\sum_{
	    \begin{subarray}{c}
		1\le q_{1},q_{2},q_{3}\le p, \\
		q_{1}\not=q_{2},\, q_{2}\not=q_{3},\, q_{3}\not=q_{1}
	    \end{subarray}
	}d_{1}^{(q_{1})}d_{2}^{(q_{2})}d_{3}^{(q_{3})}+n_{1,2,3}(v^{(p+1)})+d_{1}^{(p+1)}\Bigl(\sum_{q=1}^{p}n_{2,3}(v^{(q)})+\sum_{
		\begin{subarray}{c}
		    1\le q_{2},q_{3}\le p, \\
		    q_{2}\not=q_{3}
		\end{subarray}
	}d_{2}^{(q_{2})}d_{3}^{(q_{3})}\Bigr) \nn \\&\quad +d_{2}^{(p+1)}\Bigl(\sum_{q=1}^{p}n_{1,3}(v^{(q)})+\sum_{
		\begin{subarray}{c}
		    1\le q_{1},q_{3}\le p, \\
		    q_{1}\not=q_{3}
		\end{subarray}
	}d_{1}^{(q_{1})}d_{3}^{(q_{3})}\Bigr)+d_{3}^{(p+1)}\Bigl(\sum_{q=1}^{p}n_{1,2}(v^{(q)})+\sum_{
		\begin{subarray}{c}
		    1\le q_{1},q_{2}\le p, \\
		    q_{1}\not=q_{2}
		\end{subarray}
	}d_{1}^{(q_{1})}d_{2}^{(q_{2})}\Bigr) \nn \\
	&\quad +\Bigl(\sum_{q=1}^{p}d_{1}^{(q)}\Bigr)n_{2,3}(v^{(p+1)})+\Bigl(\sum_{q=1}^{p}d_{2}^{(q)}\Bigr)n_{1,3}(v^{(p+1)})+\Bigl(\sum_{q=1}^{p}d_{3}^{(q)}\Bigr)n_{1,2}(v^{(p+1)})+n_{1,2,3}(v^{(p+1)}) \nn \\
	&\quad =\sum_{q=1}^{p+1} n_{1,2,3}(v^{(q)}) \nn \\
	&\quad +\Bigl(\sum_{q=1}^{p+1}d_{1}^{(q)}\Bigr)\Bigl(\sum_{q=1}^{p+1}n_{2,3}(v^{(q)})\Bigr)+\Bigl(\sum_{q=1}^{p+1}d_{2}^{(q)}\Bigr)\Bigl(\sum_{q=1}^{p+1}n_{1,3}(v^{(q)})\Bigr)+\Bigl(\sum_{q=1}^{p+1}d_{3}^{(q)}\Bigr)\Bigl(\sum_{q=1}^{p+1}n_{1,2}(v^{(q)})\Bigr) \nn \\
	&\quad -\sum_{q=1}^{p+1}d_{1}^{(q)}n_{2,3}(v^{(q)})-\sum_{q=1}^{p+1}d_{2}^{(q)}n_{1,3}(v^{(q)})-\sum_{q=1}^{p+1}d_{3}^{(q)}n_{1,2}(v^{(q)}) \nn \\
	&\quad +\sum_{
	    \begin{subarray}{c}
		1\le q_{1},q_{2},q_{3}\le p+1, \\
		q_{1}\not=q_{2},\, q_{2}\not=q_{3},\, q_{3}\not=q_{1}
	    \end{subarray}
	}d_{1}^{(q_{1})}d_{2}^{(q_{2})}d_{3}^{(q_{3})} ,
    \end{align}
    which completes the proof.
\end{appendixproof}
%------------------------------------------------------------------------------

As an example of application of Theorem~\ref{direct_sum_theorem}, we choose
\begin{align}
    &V_{a}^{(1)}=\dotsb=V_{a}^{(p)}, \label{theorem_example_v_a} \\
    &n_{a}(v^{(1)})=\dotsb=n_{a}(v^{(p)}), \label{theorem_example_n_a} \\
    &n_{b,c}(v^{(1)})=\dotsb=n_{b,c}(v^{(p)}), \label{theorem_example_n_b_c} \\
    &n_{1,2,3}(v^{(1)})=\dotsb=n_{1,2,3}(v^{(p)}) \label{theorem_example_n_1_2_3}
\end{align}
and find
\begin{align}
    &n_{a}\bigl((+_{1,2,3})_{q=1}^{p}v^{(q)}\bigr)=pn_{a}(v^{(1)}), \label{theorem_equal_components_n_a} \\
    &n_{b,c}\bigl((+_{1,2,3})_{q=1}^{p}v^{(q)}\bigr)=pn_{b,c}(v^{(1)})+p(p-1)d_{b}^{(1)}d_{c}^{(1)}, \label{theorem_equal_components_n_b_c} \\
    &n_{1,2,3}\bigl((+_{1,2,3})_{q=1}^{p}v^{(q)}\bigr)=pn_{1,2,3}(v^{(1)})+p(p-1)\bigl(d_{1}^{(1)}n_{2,3}(v^{(1)})+d_{2}^{(1)}n_{1,3}(v^{(1)}) \nn \\
    &\quad +d_{3}^{(1)}n_{1,2}(v^{(1)})\bigr)+p(p-1)(p-2)d_{1}^{(1)}d_{2}^{(1)}d_{3}^{(1)}. \label{theorem_equal_components_n_1_2_3_vectors}
\end{align}
In particular, for
\begin{align}
    &p=3, \ V_{a}^{(q)}=\linearspan{\{e_{a,q}\}}, \ d_{a}^{(q)}=1, \ v^{(q)} =e_{1,q}\otimes e_{2,q}\otimes e_{3,q}, \ 1\le a\le 3, \ 1\le q\le 3, \label{}
\end{align}
we have
\begin{align}
    &n_{a}(v^{(q)})=0, \ n_{b,c}(v^{(q)})=0, \ n_{1,2,3}(v^{(q)})=0, \label{} \\
    &n_{a}\bigl((+_{1,2,3})_{q=1}^{p}v^{(q)}\bigr)=0, \ n_{b,c}\bigl((+_{1,2,3})_{q=1}^{p}v^{(q)}\bigr)=0, \ n_{1,2,3}\bigl((+_{1,2,3})_{q=1}^{p}v^{(q)}\bigr)=6, \label{}
\end{align}
for all $1\le q\le 3$, $1\le a\le 3$, $1\le b\not=c\le 3$, which coincides with the entry for the class $C_{34}$ in Table \ref{table_3_3_3}. 

%------------------------------------------------------------------------------
\subsection{Projections and annihilations}
%------------------------------------------------------------------------------

For direct sum decompositions $V_{a}=V'_{a}\oplus_{a} V''_{a}$, we define the projection operators
\begin{align}
    &P'_{a}\colon V_{a}\to V'_{a}, \ P'_{a}(v'_{a}+v''_{a})=v'_{a}, \\
    &P''_{a}\colon V_{a}\to V''_{a}, \ P''_{a}(v'_{a}+v''_{a})=v''_{a}
    \label{}
\end{align}
for any $v'_{a}\in V'_{a}$ and $v''_{a}\in V''_{a}$.
It will be easier to formulate some of our result in terms of the annihilation operators
\begin{align}
    &A'_{a}\colon V_{a}\to V_{a}, \ A'_{a}(v'_{a}+v''_{a})=\tilde{v}'_{a}, \\
    &A''_{a}\colon V_{a}\to V_{a}, \ A''_{a}(v'_{a}+v''_{a})=\tilde{v}''_{a},
    \label{}
\end{align}
where $\tilde{v}'_{a}$ and $\tilde{v}''_{a}$ are, respectively, the vectors $v'_{a}$ and $v''_{a}$ embedded into $V_{a}$.

When these operators act on tensor products of spaces, they affect only a single space and we need to indicate that the remaining two spaces are not modified by including the identity operators, e.g.~$P'_{1}\otimes\id_{2}\otimes\id_{3}\colon V_{1}\otimes V_{2}\otimes V_{3}\to V'_{1}\otimes V_{2}\otimes V_{3}$; for brevity, however, we will omit these identity operators and write simply $P'_{1}$ instead of $P'_{1}\otimes\id_{2}\otimes\id_{3}$.

Any projection of $V_{a}$ can be written as a product of irreducible projections acting on subspaces of $V_{a}$ each of which reduces the dimensionality of space by $1$.
If the one-dimensional projected space is spanned by the basis vector $e_{a,i}$, we denote the corresponding projection and annihilation by $P_{a,i}$ and $A_{a,i}$, respectively.
Both $P_{a,i}$ and $A_{a,i}$ set the $i$th component of every vector in $V_{a}$ to zero,
\begin{align}
    P_{a,i}\Bigl(\sum_{l=1}^{d_{a}}v_{a,l}e_{a,l}\Bigr)=\sum_{l=1}^{i-1}v_{a,l}e_{a,l}+\sum_{l=i+1}^{d_{a}}v_{a,l}e_{a,l}.
    \label{}
\end{align}
Any projection or annihilation of $V_{1}\otimes V_{2}\otimes V_{3}$ can be written as a product of irreducible projections $P_{1,i}$, $P_{2,j}$, $P_{3,k}$ or annihilations $A_{1,i}$, $A_{2,j}$, $A_{3,k}$, respectively.

For a given entanglement class $C$, a vector $v\in C$ and an annihilation operator $A_{a,l}$, the vector $A_{a,l}v$ belongs to a set which we denote $A_{a,l}C$.
If we consider a number $k$ of such operations for all vectors from $C$ and all operators $A_{a,l}$, we define the set
\begin{align}
    A^{k} C=\bigl\{\tilde{C}\colon A_{a_k,l_k}\dotsb A_{a_{1},l_{1}}v\in\tilde{C}, \ v\in C, \ 1\le a_{j}\le 3, \ 1\le l_{j}\le 3, \ 1\le j\le k\bigr\} \label{annihilated_classes}
\end{align}
for each $k\ge 0$ (with the convention $A^0 C=C$).
As $k$ increases, the set $A^{k} C$ becomes larger until $k=k_{\textrm{max}}$, where $k_{\textrm{max}}$ is an integer that depends on $C$, after which there are no new classes in $A^{k} C$,
\begin{align}
    &C=A^0 C\subset A^1 C\subset\dotsb\subset A^{k_{\textrm{max}}}C=A^{k_{\textrm{max}}+1}C=\dotsb.
    \label{}
\end{align}
We also define the sets $A^{k} C\setminus(\cup_{m=1}^{k-1}A^{m}C)$ such that elements of $A^{k} C\setminus(\cup_{m=1}^{k-1}A^{m}C)$ are the sets of vectors that can be obtained from $C$ by exactly $k$ applications of annihilation operators.

%------------------------------------------------------------------------------
\section{Decomposition into irreducible entanglement classes} \label{section_decomposition}
%------------------------------------------------------------------------------

The results of Section~\ref{section_direct_sum_1_2_3} naturally lead to the set of irreducible entanglement classes of $V$, which we define as follows.

Let $C$ be an entanglement class.
Suppose there does not exist a vector $v\in C$ for which there is a choice of the direct sum decompositions $V_{1}=V'_{1}\oplus_{1} V''_{1}$, $V_{2}=V'_{2}\oplus_{2} V''_{2}$, $V_{3}=V'_{3}\oplus_{3} V''_{3}$ that leads to $v=v'+v''$, where $v'\in V'$, $v''\in V''$ and $V'=V'_{1}\otimes V'_{2}\otimes V'_{3}$, $V''=V''_{1}\otimes V''_{2}\otimes V''_{3}$.
In this case we call the class $C$ irreducible.
If there exists at least one such vector for which there is at least one choice of such decompositions, then we call $C$ reducible and write, for each such choice, $C=C'\oplus_{1,2,3} C''$, where $C'$, $C''$ are the equivalence classes of $v'$, $v''$, respectively.
(Since entanglement classes are not vector spaces, the use of the direct sum symbol for classes is symbolic and it only means to remind us about the direct sum decompositions of the underlying vector spaces.)
Repeating this process a finite number of times until no further repetitions are possible (and following all possible branches due to possible multiple choices for decompositions at each step), we arrive for each $v\in V$ at one of the following representations:
\begin{align}
    &V_{a}=(\oplus_{a})_{q_{a}=1}^{p(v)} V_{a}^{(q_{a})}, \ 1\le p(v)\le d_{a}, \ 1\le a\le 3, \\
    &V^{(q)}=V_{1}^{(q)}\otimes V_{2}^{(q)}\otimes V_{3}^{(q)}, \ 1\le q\le p(v), \\
    &v=\sum_{q=1}^{p(v)} v^{(q)}, \ v^{(q)}\in V^{(q)}, \\
    &C(v)=(\oplus_{1,2,3})_{q=1}^{p(v)} C^{(q)}(v), \ v^{(q)}\in C^{(q)}(v). \label{direct_sum_of_classes}
\end{align}
Here different branches due to possible multiple choices for decompositions at each step may result in different $p(v)$, $V_{a}^{(q_{a})}$, $V^{(q)}$, $v^{(q)}$, $C^{(q)}(v)$.
We record all such resulting sets of irreducible classes $\{C^{(q)}(v)\}_{q=1}^{p(v)}$.
(Note that in view of the commutative and associative properties of the direct sum operation, as explained in Section~\ref{section_direct_sum_1_2_3}, the sum in \eqref{direct_sum_of_classes} is well-defined as long as we remember its symbolic meaning.)

We repeat the process described in the proceeding paragraph for each vector in $V$, take the union of all such found sets of irreducible classes,
\begin{align}
    &\widehat{C}(V)=\cup_{v\in V}\cup_{q=1}^{p(v)}\{C^{(q)}(v)\},
    \label{}
\end{align}
and call
\begin{align}
    \widehat{C}(V)=\{\widehat{C}^{(m)}(V)\}_{m=1}^M
    \label{irreducible_classes}
\end{align}
the set of irreducible entanglement classes of $V$. 
Once the set $\widehat{C}(V)$ is found, for each entanglement class $C$ and each vector $v\in C$, we write
\begin{align}
    &C=(\oplus_{1,2,3})_{q=1}^{p(v)} C^{(q)}, \ C^{(q)}\in\widehat{C}(V), \\
    &v=\sum_{q=1}^{p(v)} v^{(q)}, \ v^{(q)}\in C^{(q)}, \label{vector_direct_sum}
\end{align}
call $v^{(q)}$ an irreducible component of $v$ corresponding to an irreducible component $C^{(q)}$ of $C$, and view \eqref{vector_direct_sum} as a way of expressing each element of a class as a sum of elements of each of its components.

Continuing with the example considered immediately after Theorem~\ref{direct_sum_theorem}, we find that for
\begin{align}
    &V_{a}^{(1)}=\dotsb=V_{a}^{(p)}, \label{} \\
    &C^{(1)}=\dotsb=C^{(p)} \label{}
\end{align}
we have
\begin{align}
    &n_{a}\bigl((\oplus_{1,2,3})_{q=1}^{p} C^{(1)}\bigr)=pn_{a}(C^{(1)}), \label{theorem_equal_components_a}\\
    &n_{b,c}\bigl((\oplus_{1,2,3})_{q=1}^{p} C^{(1)}\bigr)=pn_{b,c}(C^{(1)})+p(p-1)d_{b}^{(1)}d_{c}^{(1)}, \label{theorem_equal_components_b_c}\\
    &n_{1,2,3}\bigl((\oplus_{1,2,3})_{q=1}^{p} C^{(1)}\bigr)=pn_{1,2,3}(C^{(1)})+p(p-1)\bigl(d_{1}^{(1)}n_{2,3}^{(1)}(C^{(1)})+d_{2}^{(1)}n_{1,3}^{(1)}(C^{(1)}) \nn \\
    &\quad +d_{3}^{(1)}n_{1,2}^{(1)}(C^{(1)})\bigr) +p(p-1)(p-2)d_{1}^{(1)}d_{2}^{(1)}d_{3}^{(1)}. \label{theorem_equal_components_1_2_3}
\end{align}

%------------------------------------------------------------------------------
\section{Examples of irreducible entanglement classes} \label{section_irreducible_classes}
%------------------------------------------------------------------------------

Since Theorem~\ref{direct_sum_theorem} allows computations of all invariants for any class $C=(\oplus_{1,2,3})_{q=1}^p C^{(q)}$ if the corresponding invariants are known for each of its (reducible or irreducible) component $C^{(q)}$, we now reduced the problem of finding invariants for any tripartite system to finding the set $\widehat{C}(V)=\{\widehat{C}^{(m)}(V)\}_{m=1}^M$ of all irreducible entanglement classes and computing the invariants for each $\widehat{C}^{(m)}(V)$.

Unfortunately we do not know how to find $\widehat{C}(V)$ for an arbitrary $V$ and restrict ourselves only to selected spaces $V$, organized mostly by their dimensions.
The rest of this section contains the statements of the results, of which only a few are with derivations.
Although these results are not very numerous, their combinations via direct sums are nevertheless sufficient together with Theorem~\ref{direct_sum_theorem} to obtain the full list of entanglement classes and their invariants for any tripartite system with dimensions $d_1\le 3$, $d_2\le 3$, $d_3\le 3$.
We give the exact correspondence for the case of three tribits in Section~\ref{section_3_3_3}. 

In the following subsections our notations for a general class in $V=V_{1,2,3}$ is $Q_c^{d_1,d_2,d_3}$, where $Q\in\{0,D,I,W,X,Y,Z\}$, $(d_1,d_2,d_3)$ are the dimensions of the spaces $(V_1,V_2,V_3)$ and $c$ is the number of nonzero terms in a representative element $v(Q_c^{d_1,d_2,d_3})$ of the class $Q_c^{d_1,d_2,d_3}$.
We note that for some classes $Q_c^{d_1,d_2,d_3}$ there exist representative elements with smaller numbers $c$, but our choice for $v(Q_c^{d_1,d_2,d_3})$ is often determined by symmetry.

The above notations allows us to enumerate a limited number of families of entanglement classes labeled by the dimensions $d_1,d_2,d_3$. 
However, as $d_1,d_2,d_3$ increase, new families of classes appear and we wish to set up a general notation for such cases.
To describe one such notation, we first introduce representing diagrams for vectors.

\begin{figure}[htpb]
    \centering
    \includegraphics[width=100pt]{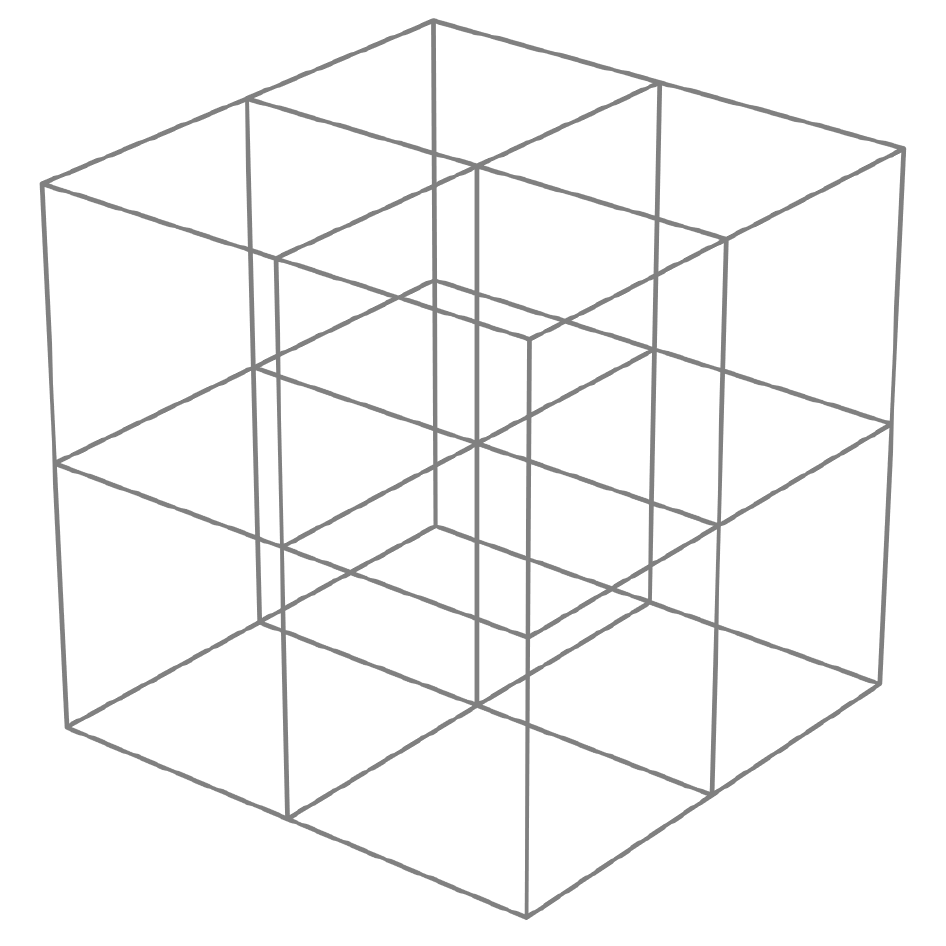}
    \caption{A diagram representing $V_1\otimes V_2\otimes V_3$ for $d_1=d_2=d_3=3$.}
    \label{figure_diagram_0}
\end{figure}

With $(e_{1,i})_{1\le i\le d_1}$, $(e_{2,j})_{1\le j\le d_2}$, $(e_{3,k})_{1\le k\le d_3}$ as respective bases for $V_1$, $V_2$, $V_3$, we represent the vector $e_{1,i}\otimes e_{2,j}\otimes e_{3,k}$ in the corresponding basis of $V_1\otimes V_2\otimes V_3$ as a lattice point $(i,j,k)$ in the three-dimensional integer lattice.
(We note that the lattice point $(i,j,k)$ has the Cartesian coordinates $(i-1,j-1,k-1)$.
    The appearance of $-1$ here and elsewhere is the consequence of the convention in which the states for each $V_a$ are labeled by $1,2,\dotsc,d_a$.
An alternative labeling by $0,1,\dotsc,d_a-1$ has its own drawbacks for the expressions of the entanglement invariants.)
We connect the lattice points by the coordinate lines and obtain the $d_1\times d_2\times d_3$ integer lattice as a representation of $V=V_1\otimes V_2\otimes V_3$; see Figure~\ref{figure_diagram_0}. 

%------------------------------------------------------------------------------
%------------------------------------------------------------------------------

It will be convenient to use such diagrams to represent vectors in $V$.
It turns out that it is sufficient for the purposes of entanglement classification to consider only vectors $v\in V$ with coordinates $v_{i,j,k}\in\{0,1\}$, and so we represent each such $v$ by an integer lattice in which all points corresponding to nonzero $v_{i,j,k}$ are marked.
(Since the normalization of $v$ does not change its entanglement invariants, all vectors that we consider are not normalized.)
To improve visualization, we connect some of these points with line segments and draw some polygons to which these points belong.
See the figures in the remainder of this section for examples of such diagrams.
Since no lattice points in the diagram in Figure~\ref{figure_diagram_0} are marked, it represents the vector $v=0$.

We now consider various entanglement classes specified by such diagrams and give the values of their invariants. 

%------------------------------------------------------------------------------
\subsection{General dimensions}
%------------------------------------------------------------------------------

%------------------------------------------------------------------------------
\subsubsection{The classes $0_{0}^{d_1,d_2,d_3}$}
%------------------------------------------------------------------------------

We define the reducible class $0_0^{d_1,d_2,d_3}$ as the equivalence class of the vector $v(0_0^{d_1,d_2,d_3})=0$, the invariants for which are trivial to compute directly and are given in Table~\ref{table_0}.
It is straightforward to construct $0_0^{d_1,d_2,d_3}$ by using $0_0^{1,1,1}$ as the building block and to verify that Theorem~\ref{direct_sum_theorem} leads to the correct values of the invariants.
Of course, there is no need to do this construction since $0_0^{d_1,d_2,d_3}$ is very simple. 
\begin{table}[h]
    \centering
    \begin{tabular}{lllll}
	\toprule
	Class             & $n_1$ & $n_2$ & $n_3$ & $n_{1,2,3}$ \\ \midrule
	$0_0^{d_1,d_2,d_3}$       & $d_1$ & $d_2$ & $d_3$ & $d_1 d_2 d_3$ \\ 
	\bottomrule
    \end{tabular}
    \caption{The invariants for the classes $0_0^{d_1,d_2,d_3}$.}
    \label{table_0}
\end{table}

\subsubsection{The classes $I_{d}^{d_{1},d_{2},d_{3}}$}

We define the irreducible class $I_1^{1,1,1}$ as the equivalence class of the vector $v(I_1^{1,1,1})=[1,1,1]$, the invariants for which are given in Table~\ref{table_i} by setting $d=1$ in any of its rows. 
Using $I_1^{1,1,1}$ as the building block, we construct for an arbitrary $d\ge 2$ the following reducible classes:
\begin{align}
    &I_d^{d,1,1}=(\oplus_1)_{i=1}^d I_1^{1,1,1}, \ v(I_d^{d,1,1})=\sum_{i=1}^d[i,1,1], \\
    &I_d^{1,d,1}=(\oplus_2)_{i=1}^d I_1^{1,1,1}, \ v(I_d^{1,d,1})=\sum_{i=1}^d[1,i,1], \\
    &I_d^{1,1,d}=(\oplus_3)_{i=1}^d I_1^{1,1,1}, \ v(I_d^{1,1,d})=\sum_{i=1}^d[1,1,i], \\
    &I_d^{d,d,1}=(\oplus_{1,2})_{i=1}^d I_1^{1,1,1}, \ v(I_d^{d,d,1})=\sum_{i=1}^d[i,i,1], \\
    &I_d^{d,1,d}=(\oplus_{1,3})_{i=1}^d I_1^{1,1,1}, \ v(I_d^{d,1,d})=\sum_{i=1}^d[i,1,i], \\
    &I_d^{1,d,d}=(\oplus_{2,3})_{i=1}^d I_1^{1,1,1}, \ v(I_d^{1,d,d})=\sum_{i=1}^d[1,i,i], \\
    &I_d^{d,d,d}=(\oplus_{1,2,3})_{i=1}^d I_1^{1,1,1}, \ v(I_d^{d,d,d})=\sum_{i=1}^d[i,i,i],
    \label{}
\end{align}
where, as above, for each class we list its generating vector and attach labels to the direct sum symbol to specify the directions in which the direct sums act.
The invariants for these classes are given in Table~\ref{table_i}, where results are separated into exact values and exact values and bounds (where applicable) from decompositions.
All decompositions results are given by Lemmas \ref{direct_sum_lemma_1}, \ref{direct_sum_lemma_1_2}, \ref{direct_sum_lemma_1_2_3} and Theorem~\ref{direct_sum_theorem} with the values of invariants $(n_{1},n_{2},n_{3},n_{1,2,3})(I_{1}^{1,1,1})=(0,0,0,0)$ that provide an initial step in recursive calculations.
We have grouped classes in Table \ref{table_i} by numbers of dimensions in which direct sums act, and classes in each group are related by permutations of indices of vector spaces.
Consequently, there is the same number of equalities and inequalities in each group as given by Lemmas \ref{direct_sum_lemma_1}, \ref{direct_sum_lemma_1_2} and \ref{direct_sum_lemma_1_2_3} and Theorem~\ref{direct_sum_theorem}. 
We in particular note that the bounds for all invariants are saturated (i.e. a bound of each inequality matches the exact value).
We will see the same property in Table \ref{table_3_3_3_verification}.

The invariants for the class $I_{d}^{d,d,d}$ also follow directly from \eqref{theorem_equal_components_a}, \eqref{theorem_equal_components_b_c} and \eqref{theorem_equal_components_1_2_3} with the values of the invariants for the class $I_{1}^{1,1,1}$ as given above.
The class $I_{2}^{2,2,2}$ is the standard GHZ class for three qubits and we call $I_d^{d,d,d}$ for an arbitrary $d\ge 3$ the generalized GHZ class.

%------------------------------------------------------------------------------
\begin{table}[htpb]
    \centering
    \begin{tabular}{ccccccccc}
	\toprule
	\multirow{2}{*}{Class} &
	\multicolumn{4}{c}{exact} &
	\multicolumn{4}{c}{from decompositions} \\
	\cmidrule(lr){2-5} \cmidrule(lr){6-9}
	& $n_1$ & $n_2$ & $n_3$ & $n_{1,2,3}$ & $n_1$ & $n_2$ & $n_3$ & $n_{1,2,3}$ \\ \midrule
	$I_{d}^{d,1,1}$ & $d-1$ & $0$   & $0$   & $0$ & $\ge d-1$ & $\ge 0$   & $\ge 0$   & $\ge 0$ \\ 
	$I_{d}^{1,d,1}$ & $0$   & $d-1$ & $0$   & $0$ & $\ge 0$   & $\ge d-1$ & $\ge 0$   & $\ge 0$ \\ 
	$I_{d}^{1,1,d}$ & $0$   & $0$   & $d-1$ & $0$ & $\ge 0$   & $\ge 0$   & $\ge d-1$ & $\ge 0$ \\ \midrule 
	$I_{d}^{d,d,1}$ & $0$   & $0$   & $0$   & $0$ & $0$   & $0$   & $\ge 0$   & $\ge 0$ \\ 
	$I_{d}^{d,1,d}$ & $0$   & $0$   & $0$   & $0$ & $0$   & $\ge 0$   & $0$   & $\ge 0$ \\ 
	$I_{d}^{1,d,d}$ & $0$   & $0$   & $0$   & $0$ & $\ge 0$   & $0$   & $0$   & $\ge 0$ \\ \midrule 
	$I_{d}^{d,d,d}$ & $0$   & $0$   & $0$   & $d(d-1)(d-2)$ & $0$   & $0$   & $0$   & $d(d-1)(d-2)$ \\ 
	\bottomrule
    \end{tabular}
    \caption{The invariants for classes constructed from the class $I_1^{1,1,1}$.
	The values and bounds in the last four columns are obtained from equalities and inequalities for the direct sum decompositions of tensor product spaces in Lemmas \ref{direct_sum_lemma_1}, \ref{direct_sum_lemma_1_2}, \ref{direct_sum_lemma_1_2_3} and Theorem \ref{direct_sum_theorem}.
    }
    \label{table_i}
\end{table}
%------------------------------------------------------------------------------

%------------------------------------------------------------------------------
\subsection{Small dimensions}
%------------------------------------------------------------------------------

\begin{figure}[htpb]
    \centering
    \includegraphics[width=125pt]{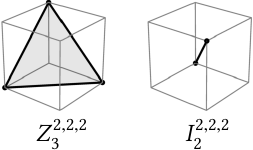}
    \caption{Diagrams of representative elements for the irreducible classes with $(d_1,d_2,d_3)=(2,2,2)$.}
    \label{figure_diagram_2_2_2}
\end{figure}

\begin{figure}[htpb]
    \centering
    \includegraphics[width=200pt]{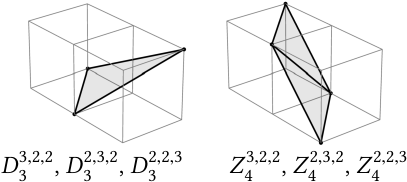}
    \caption{Diagrams of representative elements for the irreducible classes with $(d_1,d_2,d_3)\in\{(3,2,2),(2,3,2),(2,2,3)\}$.}
    \label{figure_diagram_3_2_2}
\end{figure}

\begin{figure}[htpb]
    \centering
    \includegraphics[width=350pt]{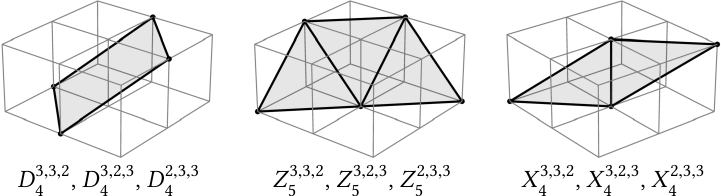}
    \caption{Diagrams of representative elements for the irreducible classes with $(d_1,d_2,d_3)\in\{(3,3,2),(3,2,3),(2,3,3)\}$.}
    \label{figure_diagram_3_3_2}
\end{figure}

\begin{figure}[htpb]
    \centering
    \includegraphics[width=350pt]{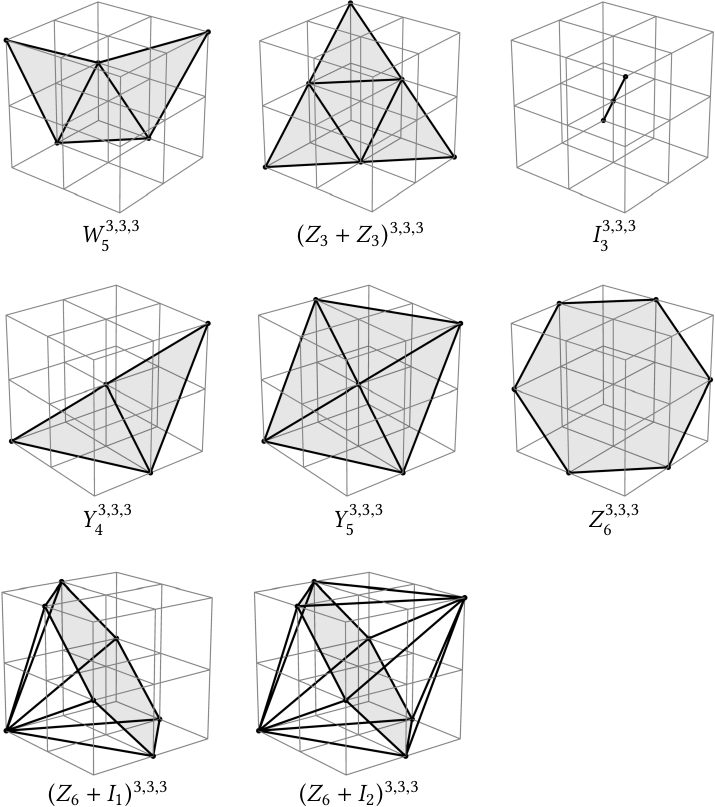}
    \caption{Diagrams of representative elements for the irreducible classes with $(d_1,d_2,d_3)=(3,3,3)$.
    Note that $Y_5^{3,3,3}\simeq Z_6^{3,3,3}$; see Table \ref{table_dimensions_3_3_3}.}
    \label{figure_diagram_3_3_3}
\end{figure}

It is convenient to group irreducible entanglement classes by the dimensions of the vector spaces.
See Figs.~\ref{figure_diagram_2_2_2}, \ref{figure_diagram_3_2_2}, \ref{figure_diagram_3_3_2}, \ref{figure_diagram_3_3_3} for diagrams of representative elements for the irreducible classes and Tables~\ref{table_dimensions_2_2_2}, \ref{table_dimensions_3_2_2}, \ref{table_dimensions_3_3_2}, \ref{table_dimensions_3_3_3} for their invariants and explicit forms of representative elements.

We have chosen to name the class $Z_{3}^{2,2,2}$ in Figure \ref{figure_diagram_2_2_2} and Table \ref{table_dimensions_2_2_2} based on the cyclic symmetry $C_{3}$ (also frequently denoted $Z_{3}$) of a symmetric form of its chosen representative element.
The same symmetry holds for the class $(Z_{3}+Z_{3})^{3,3,3}$ in Figure \ref{figure_diagram_3_3_3} and Table \ref{table_dimensions_3_3_3}.
Similarly, the chosen representative elements of $Z_{6}^{3,3,3}$, $(Z_{6}+I_{1})^{3,3,3}$ and $(Z_{6}+I_{2})^{3,3,3}$ in Figure \ref{figure_diagram_3_3_3} and Table \ref{table_dimensions_3_3_3} have the cyclic symmetry $C_{6}$ ($Z_{6}$).
We hope that the notation for the class $Z_{3}^{2,2,2}$ will not cause confusion as this class is usually called the $W$ class (especially since we also have the class $W_{5}^{3,3,3}$).

%------------------------------------------------------------------------------
%------------------------------------------------------------------------------

%------------------------------------------------------------------------------
\begin{table}[htpb]
    \centering
    \begin{tabular}{llllll}
	\toprule
	Class         & $n_1$ & $n_2$ & $n_3$ & $n_{1,2,3}$ & $v$ \\ \midrule
	$Z_3^{2,2,2}$ & $0$   & $0$   & $0$   & $1$       & $[2,1,1]+[1,2,1]+[1,1,2]$ \\ 
	$I_2^{2,2,2}$ & $0$   & $0$   & $0$   & $0$       & $[1,1,1]+[2,2,2]$ \\ 
	\bottomrule
    \end{tabular}
    \caption{The invariants and representative elements (chosen in a symmetric way) for the irreducible classes with $(d_1,d_2,d_3)=(2,2,2)$.}
    \label{table_dimensions_2_2_2}
\end{table}
%------------------------------------------------------------------------------

%------------------------------------------------------------------------------
%------------------------------------------------------------------------------

%------------------------------------------------------------------------------
\begin{table}[htpb]
    \centering
    \begin{tabular}{llllll}
	\toprule
	Class         & $n_1$ & $n_2$ & $n_3$ & $n_{1,2,3}$ & $v$ \\ \midrule
	$D_3^{3,2,2}$ & $0$   & $0$   & $0$   & $1$       & $[1,1,2]+[2,1,1]+[3,2,2]\simeq [1,2,1]+[2,1,1]+[3,2,2]$ \\ 
	$D_3^{2,3,2}$ & $0$   & $0$   & $0$   & $1$       & $[1,1,2]+[1,2,1]+[2,3,2]\simeq [1,2,1]+[2,1,1]+[2,3,2]$ \\ 
	$D_3^{2,2,3}$ & $0$   & $0$   & $0$   & $1$       & $[1,1,2]+[1,2,1]+[2,2,3]\simeq [1,1,2]+[2,1,1]+[2,2,3]$ \\ 
	$Z_4^{3,2,2}$ & $0$   & $0$   & $0$   & $0$       & $[1,2,2]+[2,1,2]+[2,2,1]+[3,1,1]$ \\ 
	$Z_4^{2,3,2}$ & $0$   & $0$   & $0$   & $0$       & $[1,2,2]+[2,1,2]+[2,2,1]+[1,3,1]$ \\ 
	$Z_4^{2,2,3}$ & $0$   & $0$   & $0$   & $0$       & $[1,2,2]+[2,1,2]+[2,2,1]+[1,1,3]$ \\
	\bottomrule
    \end{tabular}
    \caption{The invariants and representative elements for the irreducible classes with $(d_1,d_2,d_3)\in\{(3,2,2),(2,3,2),(2,2,3)\}$.}
    \label{table_dimensions_3_2_2}
\end{table}
%------------------------------------------------------------------------------

%------------------------------------------------------------------------------
%------------------------------------------------------------------------------

%------------------------------------------------------------------------------
\begin{table}[htpb]
    \centering
    \begin{tabular}{llllll}
	\toprule
	Class         & $n_1$ & $n_2$ & $n_3$ & $n_{1,2,3}$ & $v$ \\ \midrule
	$D_4^{3,3,2}$ & $0$   & $0$   & $0$   & $4$       & $[1,2,1]+[2,1,1]+[2,3,2]+[3,2,2]$ \\ 
	$D_4^{3,2,3}$ & $0$   & $0$   & $0$   & $4$       & $[1,1,2]+[2,1,1]+[2,2,3]+[3,2,2]$ \\ 
	$D_4^{2,3,3}$ & $0$   & $0$   & $0$   & $4$       & $[1,1,2]+[1,2,1]+[2,2,3]+[2,3,2]$ \\ 
	$Z_5^{3,3,2}$ & $0$   & $0$   & $0$   & $2$       & $[1,2,2]+[2,1,2]+[2,2,1]+[3,1,1]+[1,3,1]$ \\ 
	$Z_5^{3,2,3}$ & $0$   & $0$   & $0$   & $2$       & $[1,2,2]+[2,1,2]+[2,2,1]+[3,1,1]+[1,1,3]$ \\ 
	$Z_5^{2,3,3}$ & $0$   & $0$   & $0$   & $2$       & $[1,2,2]+[2,1,2]+[2,2,1]+[1,3,1]+[1,1,3]$ \\
	$X_4^{3,3,2}$ & $0$   & $0$   & $0$   & $0$       & $[1,1,1]+[2,2,1]+[2,2,2]+[3,3,2]$ \\ 
	$X_4^{3,2,3}$ & $0$   & $0$   & $0$   & $0$       & $[1,1,2]+[2,1,1]+[2,2,3]+[3,2,2]$ \\ 
	$X_4^{2,3,3}$ & $0$   & $0$   & $0$   & $0$       & $[1,1,2]+[1,2,1]+[2,2,3]+[2,3,2]$ \\ 
	\bottomrule
    \end{tabular}
    \caption{The invariants and representative elements for the irreducible classes with $(d_1,d_2,d_3)\in\{(3,3,2),(3,2,3),(2,3,3)\}$.}
    \label{table_dimensions_3_3_2}
\end{table}
%------------------------------------------------------------------------------

%------------------------------------------------------------------------------
%------------------------------------------------------------------------------

%------------------------------------------------------------------------------
\begin{table}[htpb]
    \centering
    \begin{tabular}{llllll}
	\toprule
	Class               & $n_1$ & $n_2$ & $n_3$ & $n_{1,2,3}$ & $v$ \\ \midrule
	$W_5^{3,3,3}$       & $0$   & $0$   & $0$   & $10$      & $[2,1,1]+[1,2,1]+[1,1,2]+[3,1,3]+[1,3,3]$ \\ 
	$(Z_3+Z_3)^{3,3,3}$ & $0$   & $0$   & $0$   & $8$       & $[1,2,2]+[2,1,2]+[2,2,1]+[3,1,1]+[1,3,1]+[1,1,3]$ \\
	$I_3^{3,3,3}$       & $0$   & $0$   & $0$   & $6$       & $[1,1,1]+[2,2,2]+[3,3,3]$ \\ 
	$Y_4^{3,3,3}$       & $0$   & $0$   & $0$   & $5$       & $[1,1,1]+[2,2,2]+[3,3,3]+[i,j,k]$ \\ 
	$Y_5^{3,3,3}$       & $0$   & $0$   & $0$   & $4$       & $[1,1,1]+[2,2,2]+[3,3,3]+[i',j',k']+[i'',j'',k'']$ \\ 
	$Z_6^{3,3,3}$       & $0$   & $0$   & $0$   & $4$       & $[1,2,3]+[1,3,2]+[2,1,3]+[2,3,1]+[3,1,2]+[3,2,1]$ \\ 
	$(Z_6+I_1)^{3,3,3}$ & $0$   & $0$   & $0$   & $3$       & $[1,1,1]+[1,2,3]+[1,3,2]+[2,1,3]+[2,3,1]+[3,1,2]+[3,2,1]$ \\ 
	$(Z_6+I_2)^{3,3,3}$ & $0$   & $0$   & $0$   & $2$       & $[1,1,1]+[3,3,3]+[1,2,3]+[1,3,2]+[2,1,3]+[2,3,1]+[3,1,2]+[3,2,1]$ \\ 
	\bottomrule
    \end{tabular}
    \caption{The invariants and representative elements for the irreducible classes with $(d_1,d_2,d_3)=(3,3,3)$.
	Note that $Y_5^{3,3,3}\simeq Z_6^{3,3,3}$; see Figure \ref{figure_diagram_3_3_3}.
    Here $(i,j,k)$, $(i',j',k')$, $(i'',j'',k'')$ are any permutations of $\{1,2,3\}$ such that $i'+i''=j'+j''=k'+k''=4$.}
    \label{table_dimensions_3_3_3}
\end{table}
%------------------------------------------------------------------------------

%------------------------------------------------------------------------------
\section{Three tribits} \label{section_3_3_3}
%------------------------------------------------------------------------------

Table~\ref{table_3_3_3} gives the list of all entanglement classes for three tribits, their invariants and representative vectors.
Table~\ref{table_3_3_3_decompositions} expresses the classes from Table~\ref{table_3_3_3} as decompositions of irreducible classes constructed in Section~\ref{section_irreducible_classes}.
More information on the example of three tribits can be found in \cite{Buniy:2022uwh}.

We verified that the results for entanglement classes in Tables \ref{table_3_3_3} and \ref{table_3_3_3_decompositions} that involve direct sums are consistent with the relations between invariants for direct sums in Lemmas \ref{direct_sum_lemma_1}, \ref{direct_sum_lemma_1_2}, \ref{direct_sum_lemma_1_2_3} and Theorem \ref{direct_sum_theorem}.
We collected these verifications in Table \ref{table_3_3_3_verification}.
We in particular note that the bounds for all invariants are saturated (i.e. a bound of each inequality matches the exact value), and recall that we saw the same property in Table \ref{table_i}.

%------------------------------------------------------------------------------
\begin{table}[htpb]
    \centering
    \begin{tabular}{ccc}
	\toprule
	\multirow{2}{*}{Class} &
	\multicolumn{2}{c}{$(n_1,n_2,n_3,n_{1,2,3})$} \\
	\cmidrule(lr){2-3}
	& exact & from decompositions \\ \midrule
	$C_1$ & $(2,2,2,20)$    & $(2,2,2,20)$ \\
	$C_2$ & $(1,1,2,17)$    & $(1,1,2,17)$ \\
	$C_5$ & $(0,0,2,16)$    & $(0,0,2,16)$ \\
	$C_8$ & $(1,1,1,14)$    & $(1,1,1,14)$ \\
	$C_9$ & $(1,1,1,13)$    & $(1,1,1,13)$ \\
	$C_{12}$ & $(1,1,0,12)$ & $(1,1,\ge 0,\ge 12)$ \\
	$C_{15}$ & $(1,0,0,11)$ & $(\ge 1,\ge 0,\ge 0,\ge 11)$ \\
	$C_{18}$ & $(1,1,0,11)$ & $(1,1,\ge 0,\ge 11)$ \\
	$C_{22}$ & $(1,0,0,10)$ & $(\ge 1,\ge 0,\ge 0,\ge 10)$ \\
	$C_{25}$ & $(1,0,0,9)$  & $(\ge 1,\ge 0,\ge 0,\ge 9)$ \\
	$C_{27}$ & $(0,0,1,8)$  & $(\ge 0,\ge 0,\ge 1,\ge 8)$ \\
	$C_{30}$ & $(0,0,0,7)$  & $(0,0,0,7)$ \\
	$C_{31}$ & $(0,0,1,7)$  & $(0,0,1,7)$ \\
	\bottomrule
    \end{tabular}
    \caption{Invariants for all entanglement classes for three tribits from Table \ref{table_3_3_3} that can be written as direct sums as in Table \ref{table_3_3_3_decompositions}.
	The values in the second column are from Table \ref{table_3_3_3} and the values and bounds in the third column are obtained from equalities and inequalities for the direct sum decompositions of tensor product spaces in Lemmas \ref{direct_sum_lemma_1}, \ref{direct_sum_lemma_1_2}, \ref{direct_sum_lemma_1_2_3} and Theorem \ref{direct_sum_theorem}.
    We consider only one class per each group of three classes in Table \ref{table_3_3_3} related by permutations of indices.}
    \label{table_3_3_3_verification}
\end{table}
%------------------------------------------------------------------------------

\begin{figure}[h!]
    \centering
    \includegraphics[width=\textwidth]{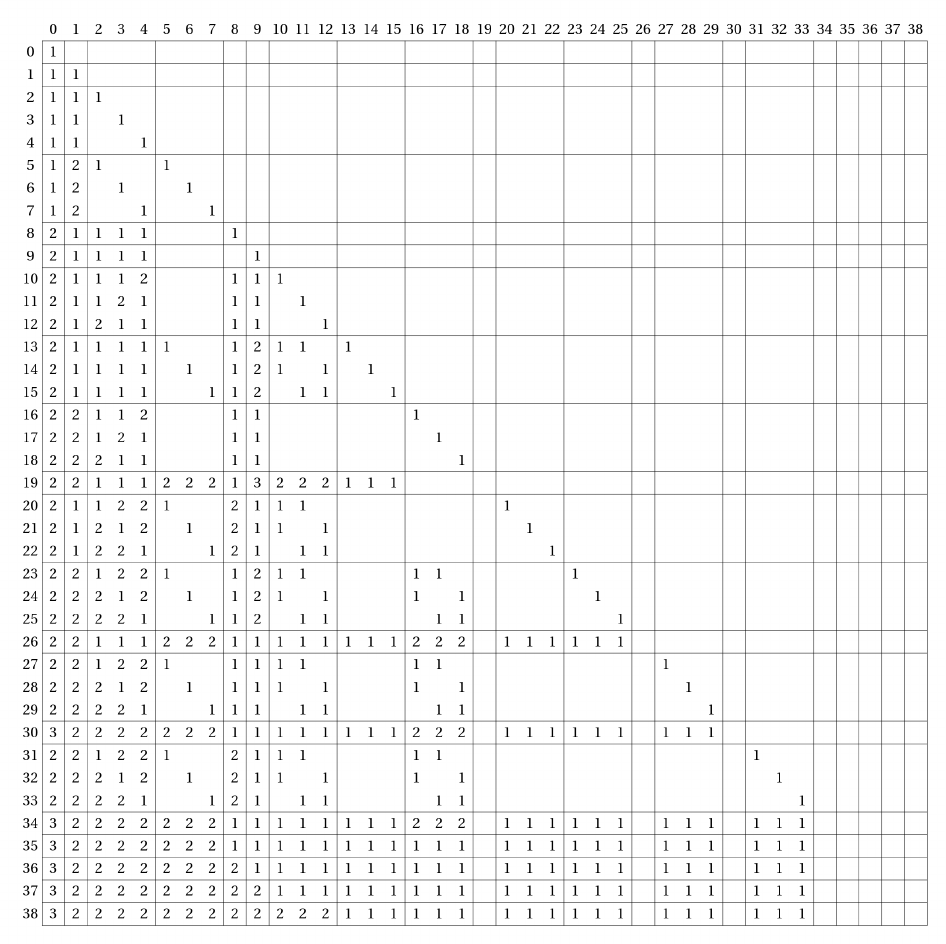}
    \caption{All the sets $A^{k} C_{i}\setminus(\cup_{m=1}^{k-1}A^{m}C_{i})$ for all the classes $C_i$ for three tribits.
	For a row number $0\le i\le 38$ and a column number $0\le j\le 38$, we have $C_j\in A^{k} C_{i}\setminus(\cup_{m=1}^{k-1}A^{m}C_{i})$, where $k$ is the number in the position $(i,j)$ in the table.
	(These $C_j$ are the only classes to which the class $C_i$ can be reduced by the number $k$ of applications of the annihilation operators $A_{a,l}$, $1\le a\le 3$, $1\le l\le 3$.)
	Only numbers $1\le k\le k_{\textrm{max}}$, where $k_{\textrm{max}}$ depends on $C_i$, are included since $A^{k} C_{i}\setminus(\cup_{m=1}^{k-1}A^{m}C_{i})=\varnothing$ for $k\ge k_{\textrm{max}}+1$. 
	If there is no number in the $(i,j)$ position, then $C_j$ does not belong to any of the sets $A^k C_i$ (and no reduction from $C_i$ to $C_j$ is possible for any $k$); equivalently, $k_{\textrm{max}}=\infty$.
    Consequently, $A^k C_i$ consists of all $C_j$ for which the label $m$ in the $(i,j)$ position is any number among $1,\dotsc,k$.}
    \label{figure_projections}
\end{figure}

Figure~\ref{figure_projections} shows all the sets $A^{k} C_{i}\setminus(\cup_{m=1}^{k-1}A^{m}C_{i})$ for all the classes $C_i$ for three tribits.
For a row number $0\le i\le 38$ and a column number $0\le j\le 38$, we have $C_j\in A^{k} C_{i}\setminus(\cup_{m=1}^{k-1}A^{m}C_{i})$, where $k$ is the number in the position $(i,j)$ in the table.
(These $C_j$ are the only classes to which the class $C_i$ can be reduced by the number $k$ of applications of the annihilation operators $A_{a,l}$, $1\le a\le 3$, $1\le l\le 3$.)
Only numbers $1\le k\le k_{\textrm{max}}$, where $k_{\textrm{max}}$ depends on $C_i$, are included since $A^{k} C_{i}\setminus(\cup_{m=1}^{k-1}A^{m}C_{i})=\varnothing$ for $k\ge k_{\textrm{max}}+1$. 
If there is no number in the $(i,j)$ position, then $C_j$ does not belong to any of the sets $A^k C_i$ (and no reduction from $C_i$ to $C_j$ is possible for any $k$); equivalently, $k_{\textrm{max}}=\infty$.
Consequently, $A^k C_i$ consists of all $C_j$ for which the label $m$ in the $(i,j)$ position is any number among $1,\dotsc,k$.

Using annihilation operators acting on vectors in the three tribits system, we can obtain the entanglement classes and representative vectors for all other tripartite systems with the dimensions satisfying $d_1\le 3$, $d_2\le 3$, $d_3\le 3$, which is equivalent to the remark we already made that the irreducible classes listed in Section~\ref{section_irreducible_classes} are sufficient to build all entanglement classes for any tripartite system with $d_1\le 3$, $d_2\le 3$, $d_3\le 3$. 

\setlength{\tabcolsep}{3pt}
\renewcommand{\arraystretch}{0.9}

%------------------------------------------------------------------------------
\begin{table}[htpb]
    \centering
    \begin{tabular}{llll}
	\toprule
	Class & $(n_1,n_2,n_3,n_{1,2,3})$ & $v$ \\ \midrule
	$C_0$ & $(3,3,3,27)$            & $0$ \\ \midrule
	$C_1$ & $(2,2,2,20)$            & $[1,1,1]$ \\ \midrule
	$C_2$ & $(1,1,2,17)$            & $[1,1,1]+[2,2,1]$ \\
	$C_3$ & $(1,2,1,17)$            & $[1,1,1]+[2,1,2]$ \\
	$C_4$ & $(2,1,1,17)$            & $[1,1,1]+[1,2,2]$ \\ \midrule
	$C_5$ & $(0,0,2,16)$            & $[1,1,1]+[2,2,1]+[3,3,1]$ \\
	$C_6$ & $(0,2,0,16)$            & $[1,1,1]+[2,1,2]+[3,1,3]$ \\
	$C_7$ & $(2,0,0,16)$            & $[1,1,1]+[1,2,2]+[1,3,3]$ \\ \midrule
	$C_8$ & $(1,1,1,14)$            & $[1,1,2]+[1,2,1]+[2,1,1]$ \\ \midrule
	$C_9$ & $(1,1,1,13)$            & $[1,1,1]+[2,2,2]$ \\ \midrule
	$C_{10}$ & $(0,1,1,12)$         & $[1,2,1]+[2,1,1]+[3,2,2]$ \\
	$C_{11}$ & $(1,0,1,12)$         & $[1,2,1]+[2,1,1]+[2,3,2]$ \\
	$C_{12}$ & $(1,1,0,12)$         & $[1,1,2]+[2,1,1]+[2,2,3]$ \\ \midrule
	$C_{13}$ & $(0,0,1,11)$         & $[1,3,2]+[2,3,1]+[3,1,2]+[3,2,1]$ \\
	$C_{14}$ & $(0,1,0,11)$         & $[1,2,3]+[2,1,3]+[3,1,2]+[3,2,1]$ \\
	$C_{15}$ & $(1,0,0,11)$         & $[1,2,3]+[1,3,2]+[2,1,3]+[2,3,1]$ \\ \midrule
	$C_{16}$ & $(0,1,1,11)$         & $[1,2,2]+[2,1,2]+[2,2,1]+[3,1,1]$ \\
	$C_{17}$ & $(1,0,1,11)$         & $[1,2,2]+[2,1,2]+[2,2,1]+[1,3,1]$ \\
	$C_{18}$ & $(1,1,0,11)$         & $[1,2,2]+[2,1,2]+[2,2,1]+[1,1,3]$ \\ \midrule
	$C_{19}$ & $(0,0,0,10)$         & $[1,1,1]+[2,1,2]+[1,2,2]+[1,3,3]+[3,3,2]$ \\ \midrule
	$C_{20}$ & $(0,0,1,10)$         & $[1,1,1]+[2,2,1]+[3,3,3]$ \\
	$C_{21}$ & $(0,1,0,10)$         & $[1,1,1]+[2,1,2]+[3,3,3]$ \\
	$C_{22}$ & $(1,0,0,10)$         & $[1,1,1]+[1,2,2]+[3,3,3]$ \\ \midrule
	$C_{23}$ & $(0,0,1,9)$          & $[1,2,2]+[2,1,2]+[2,2,1]+[1,3,1]+[3,1,1]$ \\
	$C_{24}$ & $(0,1,0,9)$          & $[1,2,2]+[2,1,2]+[2,2,1]+[1,1,3]+[3,1,1]$ \\
	$C_{25}$ & $(1,0,0,9)$          & $[1,2,2]+[2,1,2]+[2,2,1]+[1,1,3]+[1,3,1]$ \\ \midrule
	$C_{26}$ & $(0,0,0,8)$          & $[1,2,2]+[2,1,2]+[2,2,1]+[2,3,3]+[3,2,3]+[3,3,2]$ \\ \midrule
	$C_{27}$ & $(0,0,1,8)$          & $[1,1,1]+[2,2,1]+[2,3,2]+[3,2,2]$ \\
	$C_{28}$ & $(0,1,0,8)$          & $[1,1,1]+[2,1,2]+[2,2,3]+[3,2,2]$ \\
	$C_{29}$ & $(1,0,0,8)$          & $[1,1,1]+[1,2,2]+[2,2,3]+[2,3,2]$ \\ \midrule
	$C_{30}$ & $(0,0,0,7)$          & $[1,1,2]+[1,2,1]+[2,1,1]+[3,3,3]$ \\ \midrule
	$C_{31}$ & $(0,0,1,7)$          & $[1,1,1]+[2,2,1]+[2,2,2]+[3,3,2]$ \\
	$C_{32}$ & $(0,1,0,7)$          & $[1,1,1]+[2,1,2]+[2,2,2]+[3,2,3]$ \\
	$C_{33}$ & $(1,0,0,7)$          & $[1,1,1]+[1,2,2]+[2,2,2]+[2,3,3]$ \\ \midrule
	$C_{34}$ & $(0,0,0,6)$          & $[1,1,1]+[2,2,2]+[3,3,3]$ \\ \midrule
	$C_{35}$ & $(0,0,0,5)$          & $[1,1,1]+[1,2,3]+[2,2,2]+[3,3,3]$ \\ \midrule
	$C_{36}$ & $(0,0,0,4)$          & $[1,2,3]+[1,3,2]+[2,1,3]+[2,3,1]+[3,1,2]+[3,2,1]$ \\ \midrule
	$C_{37}$ & $(0,0,0,3)$          & $[1,1,1]+[1,2,3]+[1,3,2]+[2,1,3]+[2,3,1]+[3,1,2]+[3,2,1]$ \\ \midrule
	$C_{38}$ & $(0,0,0,2)$          & $[1,1,1]+[1,2,3]+[1,3,2]+[2,1,3]+[2,3,1]+[3,1,2]+[3,2,1]+[3,3,3]$ \\
	\bottomrule
    \end{tabular}
    \caption{Invariants and representative elements for all the entanglement classes for three tribits.}
    \label{table_3_3_3}
\end{table}
%------------------------------------------------------------------------------

\renewcommand{\arraystretch}{1}

%------------------------------------------------------------------------------
\begin{table}[htpb]
    \centering
    \begin{tabular}{lp{7ex}l}
	\toprule
	$C_{0}    \simeq 0_{0}^{3,3,3}$                           & & $C_{20} \simeq (I_{2}^{2,2,1}\oplus_{3} 0_{0}^{2,2,1})\oplus_{1,2,3} I_{1}^{1,1,1}$ \\ \cmidrule{1-1}
	$C_{1}    \simeq I_{1}^{1,1,1}\oplus_{1,2,3} 0_{0}^{2,2,2}$ & & $C_{21} \simeq (I_{2}^{2,1,2}\oplus_{2} 0_{0}^{2,1,2})\oplus_{1,2,3} I_{1}^{1,1,1}$ \\ \cmidrule{1-1}         
	$C_{2}    \simeq I_{2}^{2,2,1}\oplus_{1,2,3} 0_{0}^{1,1,2}$ & & $C_{22} \simeq (I_{2}^{1,2,2}\oplus_{1} 0_{0}^{1,2,2})\oplus_{1,2,3} I_{1}^{1,1,1}$ \\ \cmidrule{3-3}
	$C_{3}    \simeq I_{2}^{2,1,2}\oplus_{1,2,3} 0_{0}^{1,2,1}$ & & $C_{23} \simeq Z_{5}^{3,3,2}\oplus_{3} 0_{0}^{3,3,1}$ \\                                   
	$C_{4}    \simeq I_{2}^{1,2,2}\oplus_{1,2,3} 0_{0}^{2,1,1}$ & & $C_{24} \simeq Z_{5}^{3,2,3}\oplus_{2} 0_{0}^{3,1,3}$ \\ \cmidrule{1-1}                                  
	$C_{5}    \simeq I_{3}^{3,3,1}\oplus_{1,2,3} 0_{0}^{0,0,2}$ & & $C_{25} \simeq Z_{5}^{2,3,3}\oplus_{1} 0_{0}^{1,3,3}$ \\ \cmidrule{3-3}
	$C_{6}    \simeq I_{3}^{3,1,3}\oplus_{1,2,3} 0_{0}^{0,2,0}$ & & $C_{26} \simeq (Z_{3}+Z_{3})^{3,3,3}$ \\ \cmidrule{3-3}
	$C_{7}    \simeq I_{3}^{1,3,3}\oplus_{1,2,3} 0_{0}^{2,0,0}$ & & $C_{27} \simeq ((I_{1}^{1,1,1}\oplus_{3}0_{0}^{1,1,1})\oplus_{1,2} Z_{3}^{2,2,2})\oplus_{3} 0_{0}^{3,3,1}$ \\ \cmidrule{1-1}         
	$C_{8}    \simeq Z_{3}^{2,2,2}\oplus_{1,2,3} 0_{0}^{1,1,1}$ & & $C_{28} \simeq ((I_{1}^{1,1,1}\oplus_{2}0_{0}^{1,1,1})\oplus_{1,3} Z_{3}^{2,2,2})\oplus_{2} 0_{0}^{3,1,3}$ \\ \cmidrule{1-1}         
	$C_{9}    \simeq I_{2}^{2,2,2}\oplus_{1,2,3} 0_{0}^{1,1,1}$ & & $C_{29} \simeq ((I_{1}^{1,1,1}\oplus_{1}0_{0}^{1,1,1})\oplus_{2,3} Z_{3}^{2,2,2})\oplus_{1} 0_{0}^{1,3,3}$ \\ \cmidrule{1-1} \cmidrule{3-3}
	$C_{10} \simeq D_{3}^{3,2,2}\oplus_{2,3} 0_{0}^{3,1,1}$   & & $C_{30} \simeq I_{1}^{1,1,1}\oplus_{1,2,3} Z_{3}^{2,2,2}$ \\ \cmidrule{3-3}
	$C_{11} \simeq D_{3}^{2,3,2}\oplus_{1,3} 0_{0}^{1,3,1}$   & & $C_{31} \simeq X_{4}^{3,3,2}\oplus_{1,2,3} 0_{0}^{0,0,1}$ \\                                   
	$C_{12} \simeq D_{3}^{2,2,3}\oplus_{1,2} 0_{0}^{1,1,3}$   & & $C_{32} \simeq X_{4}^{3,2,3}\oplus_{1,2,3} 0_{0}^{0,1,0}$ \\ \cmidrule{1-1}                                   
	$C_{13} \simeq D_{4}^{3,3,2}\oplus_{3} 0_{0}^{3,3,1}$       & & $C_{33} \simeq X_{4}^{2,3,3}\oplus_{1,2,3} 0_{0}^{1,0,0}$ \\ \cmidrule{3-3}
	$C_{14} \simeq D_{4}^{3,2,3}\oplus_{2} 0_{0}^{3,1,3}$       & & $C_{34} \simeq I_{3}^{3,3,3}$ \\ \cmidrule{3-3}
	$C_{15} \simeq D_{4}^{2,3,3}\oplus_{1} 0_{0}^{1,3,3}$       & & $C_{35} \simeq Y_{4}^{3,3,3}$ \\ \cmidrule{1-1} \cmidrule{3-3}
	$C_{16} \simeq Z_{4}^{3,2,2}\oplus_{2,3} 0_{0}^{3,1,1}$   & & $C_{36} \simeq Y_{5}^{3,3,3} \simeq Z_{6}^{3,3,3}$ \\ \cmidrule{3-3}
	$C_{17} \simeq Z_{4}^{2,3,2}\oplus_{1,3} 0_{0}^{1,3,1}$   & & $C_{37} \simeq (Z_{6}+I_{1})^{3,3,3}$ \\ \cmidrule{3-3}
	$C_{18} \simeq Z_{4}^{2,2,3}\oplus_{1,2} 0_{0}^{1,1,3}$   & & $C_{38} \simeq (Z_{6}+I_{2})^{3,3,3}$ \\ \cmidrule{1-1} \cmidrule{3-3}                                               
	$C_{19} \simeq W_{5}^{3,3,3}$                           & & \\ 
	\bottomrule
    \end{tabular}
    \caption{The classes from Table~\ref{table_3_3_3} expressed in terms of the classes constructed in Section~\ref{section_irreducible_classes}.
    We attach labels to the direct sum symbols to specify unambiguously the directions in which the direct sums act.}
    \label{table_3_3_3_decompositions}
\end{table}
%------------------------------------------------------------------------------

\setlength{\tabcolsep}{3pt}

%------------------------------------------------------------------------------
%------------------------------------------------------------------------------

%------------------------------------------------------------------------------
\section{Beyond three tribits} \label{section_beyond}
%------------------------------------------------------------------------------

With the examples of irreducible classes given in Section~\ref{section_irreducible_classes} (and with any additional classes we wish to include), we can use Theorem~\ref{direct_sum_theorem} to compute the invariants for infinitely many new classes, with some or all of the dimensions $d_a>3$.
Among these, the most interesting for their entanglement properties (but only slightly more complicated to compute) are the direct sums of irreducible classes that do not have any zero diagonal blocks.
We give some of these new results in Tables~\ref{table_dimensions_1_plus_3}, \ref{table_dimensions_2_plus_2}, \ref{table_dimensions_1_plus_4}, \ref{table_dimensions_2_plus_3}, \ref{table_dimensions_1_plus_5}, \ref{table_dimensions_2_plus_4_part_1}, \ref{table_dimensions_2_plus_4_part_2}, \ref{table_dimensions_3_plus_3}. 
Since the invariants $n_1$, $n_2$, $n_3$ are all zero for all irreducible classes (which are the necessary conditions for such classes), Theorem~\ref{direct_sum_theorem} implies that these invariants remains equal to zero after the $\oplus_{1,2,3}$ operation.
For this reason, we list only the values for the invariant $n_{1,2,3}$ in these tables.
Since the invariants $n_1$, $n_2$, $n_3$, $n_{1,2,3}$ uniquely specify entanglement classes, we see there exist relations amongst the sum of blocks, e.g., from Tables~\ref{table_dimensions_1_plus_3}, \ref{table_dimensions_2_plus_2} we find
\begin{align}
    &I_{1}^{1,1,1}\oplus_{1,2,3}(Z_{3}+Z_{3})^{3,3,3}\simeq Z_{3}^{2,2,2}\oplus_{1,2,3}Z_{3}^{2,2,2},
    \label{}
\end{align}
from Tables~\ref{table_dimensions_1_plus_4}, \ref{table_dimensions_2_plus_3} that
\begin{align}
    &I_{2}^{2,2,2}\oplus_{1,2,3}(Z_{3}+Z_{3})^{3,3,3}\simeq Z_{3}^{2,2,2}\oplus_{1,2,3}Y_{4}^{3,3,3}\simeq I_{5}^{5,5,5}, \\
    &Z_{3}^{2,2,2}\oplus_{1,2,3}Z_{6}^{3,3,3}\simeq I_{2}^{2,2,2}\oplus_{1,2,3}Y_{4}^{3,3,3}, \\
    &Z_{3}^{2,2,2}\oplus_{1,2,3}(Z_{6}+I_{1})^{3,3,3}\simeq I_{2}^{2,2,2}\oplus_{1,2,3}Z_{6}^{3,3,3}, \\
    &Z_{3}^{2,2,2}\oplus_{1,2,3}(Z_{6}+I_{2})^{3,3,3}\simeq I_{2}^{2,2,2}\oplus_{1,2,3}(Z_{6}+I_{1})^{3,3,3},
    \label{}
\end{align}
and from Tables~\ref{table_dimensions_1_plus_5}, \ref{table_dimensions_2_plus_4_part_1}, \ref{table_dimensions_2_plus_4_part_2}, \ref{table_dimensions_3_plus_3} that
\begin{align}
    &W_{5}^{3,3,3}\oplus_{1,2,3}I_{3}^{3,3,3}\simeq(Z_{3}+Z_{3})^{3,3,3}\oplus_{1,2,3}(Z_{3}+Z_{3})^{3,3,3}, \\
    &I_{2}^{2,2,2}\oplus_{1,2,3}Z_{3}^{2,2,2}\oplus_{1,2,3}Z_{3}^{2,2,2}\simeq W_{5}^{3,3,3}\oplus_{1,2,3}Z_{6}^{3,3,3}\simeq I_{3}^{3,3,3}\oplus_{1,2,3}(Z_{3}+Z_{3})^{3,3,3}, \\
    &Z_{3}^{2,2,2}\oplus_{1,2,3} I_{4}^{4,4,4}\simeq W_{5}^{3,3,3}\oplus_{1,2,3}(Z_{6}+I_{1})^{3,3,3}\simeq(Z_{3}+Z_{3})^{3,3,3}\oplus_{1,2,3}Y_{4}^{3,3,3}, \\
    &W_{5}^{3,3,3}\oplus_{1,2,3}(Z_{6}+I_{2})^{3,3,3}\simeq(Z_{3}+Z_{3})^{3,3,3}\oplus_{1,2,3}Z_{6}^{3,3,3}\simeq I_{6}^{6,6,6}, \\
    &(Z_{3}+Z_{3})^{3,3,3}\oplus_{1,2,3}(Z_{6}+I_{1})^{3,3,3}\simeq I_{3}^{3,3,3}\oplus_{1,2,3} Y_{4}^{3,3,3}, \\
    &(Z_{3}+Z_{3})^{3,3,3}\oplus_{1,2,3}(Z_{6}+I_{2})^{3,3,3}\simeq I_{3}^{3,3,3}\oplus_{1,2,3} Z_{6}^{3,3,3}\simeq Y_{4}^{3,3,3}\oplus_{1,2,3} Y_{4}^{3,3,3}, \\
    &I_{3}^{3,3,3}\oplus_{1,2,3}(Z_{6}+I_{1})^{3,3,3}\simeq Y_{4}^{3,3,3}\oplus_{1,2,3} Z_{6}^{3,3,3}, \\
    &I_{3}^{3,3,3}\oplus_{1,2,3}(Z_{6}+I_{2})^{3,3,3}\simeq Y_{4}^{3,3,3}\oplus_{1,2,3}(Z_{6}+I_{1})^{3,3,3}\simeq Z_{6}^{3,3,3}\oplus_{1,2,3}Z_{6}^{3,3,3}, \\
    &Y_{4}^{3,3,3}\oplus_{1,2,3}(Z_{6}+I_{2})^{3,3,3}\simeq Z_{6}^{3,3,3}\oplus_{1,2,3}(Z_{6}+I_{1})^{3,3,3}, \\
    &Z_{6}^{3,3,3}\oplus_{1,2,3}(Z_{6}+I_{2})^{3,3,3}\simeq(Z_{6}+I_{1})^{3,3,3}\oplus_{1,2,3}(Z_{6}+I_{1})^{3,3,3}.
    \label{}
\end{align}

%------------------------------------------------------------------------------
\begin{table}[htpb]
    \centering
    \begin{tabular}{llllllll}
	\toprule
	& $W_5^{3,3,3}$ & $(Z_3+Z_3)^{3,3,3}$ & $I_3^{3,3,3}$ & $Y_4^{3,3,3}$ & $Z_6^{3,3,3}$ & $(Z_6+I_1)^{3,3,3}$ & $(Z_6+I_2)^{3,3,3}$ \\ \midrule 
	$I_1^{1,1,1}$ & $28$          & $26$                & $24$          & $23$          & $22$          & $21$                & $20$ \\
	\bottomrule
    \end{tabular}
    \caption{The invariants $n_{1,2,3}(C'\oplus_{1,2,3}C'')$ for $d'_1=d'_2=d'_3=1$, $d''_1=d''_2=d''_3=3$ and several choices of $C'$ and $C''$ computed by using Theorem \ref{direct_sum_theorem}.}
    \label{table_dimensions_1_plus_3}
\end{table}
%------------------------------------------------------------------------------

%------------------------------------------------------------------------------
\begin{table}[htpb]
    \centering
    \begin{tabular}{lll}
	\toprule
	& $Z_3^{2,2,2}$ & $I_2^{2,2,2}$ \\ \midrule 
	$Z_3^{2,2,2}$ & $26$          & $25$          \\
	$I_2^{2,2,2}$ & $25$          & $24$          \\
	\bottomrule
    \end{tabular}
    \caption{The invariants $n_{1,2,3}(C'\oplus_{1,2,3}C'')$ for $d'_1=d'_2=d'_3=2$, $d''_1=d''_2=d''_3=2$ and several choices of $C'$ and $C''$ computed by using Theorem \ref{direct_sum_theorem}.}
    \label{table_dimensions_2_plus_2}
\end{table}
%------------------------------------------------------------------------------

%------------------------------------------------------------------------------
\begin{table}[htpb]
    \centering
    \begin{tabular}{llll}
	\toprule
	& $Z_3^{2,2,2}\oplus_{1,2,3}Z_3^{2,2,2}$ & $Z_3^{2,2,2}\oplus_{1,2,3}I_2^{2,2,2}$ & $I_4^{4,4,4}$ \\ \midrule
	$I_1^{1,1,1}$ & $62$                                 & $61$                                 & $60$ \\
	\bottomrule
    \end{tabular}
    \caption{The invariants $n_{1,2,3}(C'\oplus_{1,2,3}C'')$ for $d'_1=d'_2=d'_3=1$, $d''_1=d''_2=d''_3=4$ and several choices of $C'$ and $C''$ computed by using Theorem \ref{direct_sum_theorem}.}
    \label{table_dimensions_1_plus_4}
\end{table}
%------------------------------------------------------------------------------

%------------------------------------------------------------------------------
\begin{table}[htpb]
    \centering
    \begin{tabular}{llllllll}
	\toprule
	& $W_5^{3,3,3}$ & $(Z_3+Z_3)^{3,3,3}$ & $I_3^{3,3,3}$ & $Y_4^{3,3,3}$ & $Z_6^{3,3,3}$ & $(Z_6+I_1)^{3,3,3}$ & $(Z_6+I_2)^{3,3,3}$ \\ \midrule
	$Z_3^{2,2,2}$ & $65$          & $63$                & $61$          & $60$          & $59$          & $58$                & $57$ \\
	$I_2^{2,2,2}$ & $64$          & $62$                & $60$          & $59$          & $58$          & $57$                & $56$ \\
	\bottomrule
    \end{tabular}
    \caption{The invariants $n_{1,2,3}(C'\oplus_{1,2,3}C'')$ for $d'_1=d'_2=d'_3=2$, $d''_1=d''_2=d''_3=3$ and several choices of $C'$ and $C''$ computed by using Theorem \ref{direct_sum_theorem}.}
    \label{table_dimensions_2_plus_3}
\end{table}
%------------------------------------------------------------------------------

%------------------------------------------------------------------------------
\begin{table}[htpb]
    \centering
    \begin{tabular}{llll}
	\toprule
	& $I_1^{1,1,1}\oplus_{1,2,3}Z_3^{2,2,2}\oplus_{1,2,3}Z_3^{2,2,2}$ & $Z_3^{2,2,2}\oplus_{1,2,3}I_3^{3,3,3}$ & $I_5^{5,5,5}$ \\ \midrule
	$I_1^{1,1,1}$ & $122$                                                       & $121$                                & $120$ \\
	\bottomrule
    \end{tabular}
    \caption{The invariants $n_{1,2,3}(C'\oplus_{1,2,3}C'')$ for $d'_1=d'_2=d'_3=1$, $d''_1=d''_2=d''_3=5$ and several choices of $C'$ and $C''$ computed by using Theorem \ref{direct_sum_theorem}.}
    \label{table_dimensions_1_plus_5}
\end{table}
%------------------------------------------------------------------------------

%------------------------------------------------------------------------------
\begin{table}[htpb]
    \centering
    \begin{tabular}{llllll}
	\toprule
	& $I_1^{1,1,1}\oplus_{1,2,3}W_5^{3,3,3}$ & $I_1^{1,1,1}\oplus_{1,2,3}(Z_3+Z_3)^{3,3,3}$ & $Z_3^{2,2,2}\oplus_{1,2,3}Z_3^{2,2,2}$ & $Z_3^{2,2,2}\oplus_{1,2,3} I_2^{2,2,2}$ & $I_4^{4,4,4}$ \\ \midrule
	$I_2^{2,2,2}$ & $124$                                & $122$                                      & $122$                                & $121$                                 & $120$ \\
	\bottomrule
    \end{tabular}
    \caption{The invariants $n_{1,2,3}(C'\oplus_{1,2,3}C'')$ for $d'_1=d'_2=d'_3=2$, $d''_1=d''_2=d''_3=4$ and several choices of $C'$ and $C''$ computed by using Theorem \ref{direct_sum_theorem}.}
    \label{table_dimensions_2_plus_4_part_1}
\end{table}
%------------------------------------------------------------------------------

%------------------------------------------------------------------------------
\begin{table}[htpb]
    \centering
    \begin{tabular}{lllll}
	\toprule
	& $I_1^{1,1,1}\oplus_{1,2,3}Y_4^{3,3,3}$ & $I_1^{1,1,1}\oplus_{1,2,3}Z_6^{3,3,3}$ & $I_1^{1,1,1}\oplus_{1,2,3}(Z_6+I_1)^{3,3,3}$ & $I_1^{1,1,1}\oplus_{1,2,3}(Z_6+I_2)^{3,3,3}$ \\ \midrule
	$I_2^{2,2,2}$ & $119$                                & $118$                                & $117$                                      & $116$ \\
	\bottomrule
    \end{tabular}
    \caption{The invariants $n_{1,2,3}(C'\oplus_{1,2,3}C'')$ for $d'_1=d'_2=d'_3=2$, $d''_1=d''_2=d''_3=4$ and several choices of $C'$ and $C''$ computed by using Theorem \ref{direct_sum_theorem}.}
    \label{table_dimensions_2_plus_4_part_2}
\end{table}
%------------------------------------------------------------------------------

%------------------------------------------------------------------------------
\begin{table}[htpb]
    \centering
    \begin{tabular}{llllllll}
	\toprule
	& $W_5^{3,3,3}$ & $(Z_3+Z_3)^{3,3,3}$ & $I_3^{3,3,3}$ & $Y_4^{3,3,3}$ & $Z_6^{3,3,3}$ & $(Z_6+I_1)^{3,3,3}$ & $(Z_6+I_2)^{3,3,3}$ \\ \midrule
	$W_5^{3,3,3}$       & $128$         & $126$               & $124$         & $123$         & $122$         & $121$               & $120$ \\
	$(Z_3+Z_3)^{3,3,3}$ & $126$         & $124$               & $122$         & $121$         & $120$         & $119$               & $118$ \\
	$I_3^{3,3,3}$       & $124$         & $122$               & $120$         & $119$         & $118$         & $117$               & $116$ \\
	$Y_4^{3,3,3}$       & $123$         & $121$               & $119$         & $118$         & $117$         & $116$               & $115$ \\
	$Z_6^{3,3,3}$       & $122$         & $120$               & $118$         & $117$         & $116$         & $115$               & $114$ \\
	$(Z_6+I_1)^{3,3,3}$ & $121$         & $119$               & $117$         & $116$         & $115$         & $114$               & $113$ \\
	$(Z_6+I_2)^{3,3,3}$ & $120$         & $118$               & $116$         & $115$         & $114$         & $113$               & $112$ \\
	\bottomrule
    \end{tabular}
    \caption{The invariants $n_{1,2,3}(C'\oplus_{1,2,3}C'')$ for $d'_1=d'_2=d'_3=3$, $d''_1=d''_2=d''_3=3$ and several choices of $C'$ and $C''$ computed by using Theorem \ref{direct_sum_theorem}.}
    \label{table_dimensions_3_plus_3}
\end{table}
%------------------------------------------------------------------------------

We note that the above relations say that in general there are multiple ways to decompose a given entanglement class into irreducible components.
It should be clear how to proceed with deriving similar relations by using Theorem~\ref{direct_sum_theorem}.

%------------------------------------------------------------------------------
\section{Summary and outlook} \label{section_summary}
%------------------------------------------------------------------------------

Complete classification of all entanglement states in any quantum system is presently an unsolved problem.
Instead, we proposed in \cite{Buniy:2010yh,Buniy:2010zp} a solution to the partial classification problem, where the set of all entangled states for a given system is partitioned into equivalence classes based on the values of algebraic entanglement invariants for the states.
These invariants are properties of certain linear maps associated with entangled states.
The proposed general method works for any number and dimensions of subsystems of any given system.
We applied our solution in \cite{Buniy:2010yh,Buniy:2010zp} to studying various examples of three subsystems as well as the system of four qubits.
In each case that we considered, we obtained either a new (if it was previously unknown), the same or a more refined entanglement classification.
A large majority of the results were new. 

As our studies in \cite{Buniy:2010yh,Buniy:2010zp} were focused on the development of the general method followed by explorations of numerous examples to show how the method works in practice, we did not study in detail various properties of our entanglement measures and associated entanglement classifications.
This is precisely the task that we have undertaken here and from a different perspective in \cite{Buniy:2022uwh}.
We have used several operations to build new entangled states from old entangled states (tensor products, direct sums and projections) to gain insight into relationships between entanglement invariants of states related by these operations.
This naturally led us to the general decomposition theorem (Theorem \ref{direct_sum_theorem}) and the notion of reducible and irreducible entanglement classes.
The irreducible classes are the elementary building blocks from which all other entanglement classes can be constructed with the help of the operations mentioned above.
Importantly, the theorem gives simple expressions for the algebraic invariants of any class in terms of the invariants of its components.

The decomposition theorem reduces our partial entanglement classification problem for any given system to three separate simpler problems: (i) finding all irreducible classes for a given system, (ii) finding all possible combinations of irreducible classes and (iii) computing the invariants for each of these combinations.
The theorem completely solves the third problem.
Once the first problem is solved for any given system, the found irreducible classes will also be among irreducible classes of larger systems, but the difficulty of the first problem lies in the ever-expanding (with the system's dimensions) list of irreducible classes not present in any lower-dimensional systems.
The difficulty of the second problem is combinatorial as it requires knowledge of the full list of ways of combining irreducible classes, and the allowed combinations depend on the dimensions of a system and its subsystems.

We have completely solved the first and second problem listed above for three tribits.
During these solutions we encountered a problem that is typical for studies of equivalence classes: which representative vectors do we choose for each entanglement class?
Although any such choice works (after all, these classes are called equivalence classes for a reason) it might be beneficial to select representative elements based on some general principle.
We proceeded with symmetry as our guiding principle; namely, we selected representative vectors having the smallest number of nonzero components and the highest degree of symmetry that we could find, although occasionally compromises were needed to accommodate these two requirements.
Additionally, we were guided in such selections by increases and decreases of symmetry by means of the operations which we used to construct new entangled states from old entangled states (tensor products, direct sums and projections).

Many problems remain unsolved even when one is limited to the restricted entanglement classification.
For example, although we have established the complete set of allowed values for the invariants $n_1$, $n_2$, $n_3$, we as yet have no similar result for the invariant $n_{1,2,3}$ and instead only have lower and upper bounds for it.
Another interesting problem is to find the number of entanglement classes for an arbitrary given system.
Although we have already found these numbers for numerous tripartite systems in \cite{Buniy:2010yh,Buniy:2010zp}, it would be interesting to establish the general expression for any tripartite system.
A more restricted problem of finding the asymptotic behavior of the number of classes for large dimensions seems to be doable; numerical evidence suggests that the number grows exponentially with the dimensions.
The problems (i) and (ii) mentioned above are other general unsolved problems.
We think their solutions can provide deep insights into geometry and topology of entangled states.

For example, there is an analogy between three qubit entanglement states and the topology linking of three classical rings in three dimensions \cite{Kauffman}.
In the Borromean ring configuration the three rings are linked, but if one is cut all three become unlinked.
That is similar to the entangled GHZ quantum state.
Classical rings can be pairwise entangled, but there is a topological obstruction to having both pairwise and tripartite linking of rings.
This is where the analogy breaks down as W-states can have both pairwise and tripartite entanglement, but the analogy can help with the appreciation on what true tripartite entanglement means and how it can be used.

Tripartite entanglement has been a proving ground for testing both theoretical and experimental ideas of multi-partite entanglement.
It is sufficiently complex to carry with it many of the properties required in development of concepts and devices needed for quantum information processing without being so cumbersome that one becomes bogged down in tedious calculations.
This is why we have focused on gaining a full understanding of the allowed entanglements of tripartite systems of three arbitrary qudits, knowledge of which we hope will speed development of a broad spectrum of quantum devices, including quantum computers.

%--------------------------------------------------------------------------------
\appendix
%--------------------------------------------------------------------------------

%------------------------------------------------------------------------------

%------------------------------------------------------------------------------


\begin{thebibliography}{100}

%\cite{Horodecki:2009zz}
    \bibitem{Horodecki:2009zz} 
	R.~Horodecki, P.~Horodecki, M.~Horodecki and K.~Horodecki,
  %``Quantum entanglement,''
	Rev.\ Mod.\ Phys.\  {\bf 81}, 865 (2009);
	doi:10.1103/RevModPhys.81.865;
	[quant-ph/0702225].
  %%CITATION = doi:10.1103/RevModPhys.81.865;%%
  %287 citations counted in INSPIRE as of 28 Oct 2017

%\cite{Steane:1997kb}
    \bibitem{Steane:1997kb} 
	A.~Steane,
  %``Quantum computing,''
	Rept.\ Prog.\ Phys.\  {\bf 61}, 117 (1998);
	doi:10.1088/0034-4885/61/2/002;
	[quant-ph/9708022].
  %%CITATION = doi:10.1088/0034-4885/61/2/002;%%
  %91 citations counted in INSPIRE as of 28 Oct 2017

%\cite{PreskillLecture}
    \bibitem{PreskillLecture}
	J.~Preskill, Lecture Notes: Quantum Information and Computation (1998).

%\cite{Kitaev:1997wr}
    \bibitem{Kitaev:1997wr} 
	A.~Y.~Kitaev,
  %``Fault tolerant quantum computation by anyons,''
	Annals Phys.\  {\bf 303}, 2 (2003)
	doi:10.1016/S0003-4916(02)00018-0
	[quant-ph/9707021].
  %%CITATION = doi:10.1016/S0003-4916(02)00018-0;%%
  %850 citations counted in INSPIRE as of 28 Oct 2017

%\cite{Nayak:2008zza}
    \bibitem{Nayak:2008zza} 
	C.~Nayak, S.~H.~Simon, A.~Stern, M.~Freedman and S.~Das Sarma,
  %``Non-Abelian anyons and topological quantum computation,''
	Rev.\ Mod.\ Phys.\  {\bf 80}, 1083 (2008).
	doi:10.1103/RevModPhys.80.1083
  %%CITATION = doi:10.1103/RevModPhys.80.1083;%%
  %840 citations counted in INSPIRE as of 28 Oct 2017

%\cite{NielsenChuang}
    \bibitem{NielsenChuang}
	M.~A.~Nielsen and I.~L.~Chuang
	``Quantum Computation and Quantum Information ,''
	Cambridge University Press (2000) Cambridge, UK and New York, NY, USA

%\cite{Castelvecchi}
    \bibitem{Castelvecchi}
  %``Quantum computers ready to leap out of the lab in 2017,''
	D.~Castelvecchi, Nature, 541, Issue 7635, 9 (2017);
	doi:10.1038/541009a.

%\cite{Buniy:2010yh}
    \bibitem{Buniy:2010yh} 
	R.~V.~Buniy and T.~W.~Kephart,
  %``New Invariants for Entangled States,''
	J.\ Phys.\ A {\bf 45}, 182001 (2012);
	doi:10.1088/1751-8113/45/18/182001;
	[arXiv:1009.2217 [quant-ph]].
  %%CITATION = doi:10.1088/1751-8113/45/18/182001;%%
  %1 citations counted in INSPIRE as of 28 Oct 2017

%\cite{Buniy:2010zp}
    \bibitem{Buniy:2010zp} 
	R.~V.~Buniy and T.~W.~Kephart,
  %``An Algebraic Classification of Entangled States,''
	J.\ Phys.\ A {\bf 45}, 185304 (2012);
	doi:10.1088/1751-8113/45/18/185304;
	[arXiv:1012.2630 [quant-ph]].
  %%CITATION = doi:10.1088/1751-8113/45/18/185304;%%
  %3 citations counted in INSPIRE as of 28 Oct 2017

%\cite{Neeley}
    \bibitem{Neeley} 
  %``Emulation of a quantum spin with a superconducting phase qudit,''
	M.~Neeley, et al., Science 07 Aug 2009:
	Vol. 325, Issue 5941, pp. 722-725;
	doi:10.1126/science.1173440 

%\cite{Lima}
    \bibitem{Lima} 
  %``Manipulating spatial qudit states with programmable optical devices,''
	G.~Lima, A.~Vargas, L.~Neves, R.~Guzm\'{a}n, C.~Saavedra,  
	Optics Express 17(13) 10688 (2009).

%\cite{Srivastava}
    \bibitem{Srivastava}
  %``Modelling microtubules in the brain as n-qudit quantum Hopfield network and beyond,''
	D.~P.~Srivastava, V.~Sahni and P.~S.~Satsangi, Intern. J. of Gen. Sys. 45, 41 (2016);
	arXiv:1505.00774; doi.org/10.1080/03081079.2015.1076405.

%\cite{Jafarizadeh}
    \bibitem{Jafarizadeh}
  %``An explicit formula for the polynomial entanglement measures of degree 2 of even-N qubits mixed states,''
	M.~A.~Jafarizadeh, et al., Eur. Phys. J. D 71, 254 (2017); arXiv:1611.07958 

%\cite{Wang}
    \bibitem{Wang}
  %``Trace distance measure of coherence for a class of qudit states,''
	Z.~Wang, Y.-L.~Wang, Z.-X.~Wang, Quantum Inf. Process (2016) 15: 4641;
	doi:10.1007/s11128-016-1403-z;
	arXiv:1610.07330.

%\cite{Rungta}
    \bibitem{Rungta}
  %``Qudit Entanglement,'' 
	P.~Rungta, et al., pp. 149--164, in ``Directions in Quantum Optics,'' H.~J.~Carmichael, R.~J.~Glauber, M.~O.~Scully (Eds.): LNP 561, Springer-Verlag Berlin Heidelberg 2001.

%\cite{Jamiolkowski}
    \bibitem{Jamiolkowski}
  %``Algebraic and geometric representations of the two-qudit entanglement,''
	A.~Jamiolkowski, Math. Analysis Lab. Rec., 1350, 100 (2004).

  %\cite{Ho:2016gki}
    \bibitem{Ho:2016gki}
	C.~L.~Ho and T.~Deguchi,
%``Multi-qudit states generated by unitary braid quantum gates based on Temperley-Lieb algebra,''
	EPL \textbf{118}, no.4, 40001 (2017)
	doi:10.1209/0295-5075/118/40001
	[arXiv:1611.06772 [quant-ph]].
  %3 citations counted in INSPIRE as of 26 Nov 2022

%\cite{Duff:2006ue}
    \bibitem{Duff:2006ue} 
	M.~J.~Duff and S.~Ferrara,
  %``E(7) and the tripartite entanglement of seven qubits,''
	Phys.\ Rev.\ D {\bf 76}, 025018 (2007)
	doi:10.1103/PhysRevD.76.025018
	[quant-ph/0609227].
  %%CITATION = doi:10.1103/PhysRevD.76.025018;%%
  %60 citations counted in INSPIRE as of 28 Oct 2017

%\cite{Levay:2006pt}
    \bibitem{Levay:2006pt} 
	P.~Levay,
  %``Strings, black holes, the tripartite entanglement of seven qubits and the Fano plane,''
	Phys.\ Rev.\ D {\bf 75}, 024024 (2007)
	doi:10.1103/PhysRevD.75.024024
	[hep-th/0610314].
  %%CITATION = doi:10.1103/PhysRevD.75.024024;%%
  %50 citations counted in INSPIRE as of 28 Oct 2017

%\cite{Borsten:2008wd}
    \bibitem{Borsten:2008wd} 
	L.~Borsten, D.~Dahanayake, M.~J.~Duff, H.~Ebrahim and W.~Rubens,
  %``Black Holes, Qubits and Octonions,''
	Phys.\ Rept.\  {\bf 471}, 113 (2009)
	doi:10.1016/j.physrep.2008.11.002
	[arXiv:0809.4685 [hep-th]].
  %%CITATION = doi:10.1016/j.physrep.2008.11.002;%%
  %96 citations counted in INSPIRE as of 28 Oct 2017

%\cite{Borsten:2010db}
    \bibitem{Borsten:2010db} 
	L.~Borsten, D.~Dahanayake, M.~J.~Duff, A.~Marrani and W.~Rubens,
  %``Four-qubit entanglement from string theory,''
	Phys.\ Rev.\ Lett.\  {\bf 105}, 100507 (2010)
	doi:10.1103/PhysRevLett.105.100507
	[arXiv:1005.4915 [hep-th]].
  %%CITATION = doi:10.1103/PhysRevLett.105.100507;%%
  %80 citations counted in INSPIRE as of 28 Oct 2017

%\cite{Borsten:2012fx}
    \bibitem{Borsten:2012fx} 
	L.~Borsten, M.~J.~Duff and P.~Levay,
  %``The black-hole/qubit correspondence: an up-to-date review,''
	Class.\ Quant.\ Grav.\  {\bf 29}, 224008 (2012)
	doi:10.1088/0264-9381/29/22/224008
	[arXiv:1206.3166 [hep-th]].
  %%CITATION = doi:10.1088/0264-9381/29/22/224008;%%
  %49 citations counted in INSPIRE as of 28 Oct 2017

%\cite{Levay:2006kf}
    \bibitem{Levay:2006kf} 
	P.~Levay,
  %``Stringy black holes and the geometry of entanglement,''
	Phys.\ Rev.\ D {\bf 74}, 024030 (2006)
	doi:10.1103/PhysRevD.74.024030
	[hep-th/0603136].
  %%CITATION = doi:10.1103/PhysRevD.74.024030;%%
  %59 citations counted in INSPIRE as of 28 Oct 2017

%\cite{Bina}
    \bibitem{Bina} 
  %``Tripartite quantum state mapping and discontinuous entanglement transfer in a cavity QED open system,''
	M.~Bina, et al., Physica Scripta 2010 (T140), 014015.

%\cite{Kim:2015dbb}
    \bibitem{Kim:2015dbb} 
	K.~I.~Kim, H.~M.~Li and B.~K.~Zhao,
  %``Genuine Tripartite Entanglement Dynamics and Transfer in a Triple Jaynes-Cummings Model,''
	Int.\ J.\ Theor.\ Phys.\  {\bf 55}, no. 1, 241 (2016).
	doi:10.1007/s10773-015-2656-5
  %%CITATION = doi:10.1007/s10773-015-2656-5;%%

%\cite{Hwang:2010ib}
    \bibitem{Hwang:2010ib} 
	M.~R.~Hwang, D.~Park and E.~Jung,
  %``Tripartite Entanglement in Noninertial Frame,''
	Phys.\ Rev.\ A {\bf 83}, 012111 (2001)
	doi:10.1103/PhysRevA.83.012111
	[arXiv:1010.6154 [hep-th]].
  %%CITATION = doi:10.1103/PhysRevA.83.012111;%%
  %15 citations counted in INSPIRE as of 26 Jul 2018

%\cite{Shamirzai:2011gk}
    \bibitem{Shamirzai:2011gk} 
	M.~Shamirzai, B.~N.~Esfahani and M.~Soltani,
  %``Tripartite Entanglements in Non-inertial Frames,''
	Int.\ J.\ Theor.\ Phys.\  {\bf 51}, 787 (2012)
	doi:10.1007/s10773-011-0958-9
	[arXiv:1103.0258 [quant-ph]].
  %%CITATION = doi:10.1007/s10773-011-0958-9;%%
  %3 citations counted in INSPIRE as of 30 Oct 2017

%\cite{Khan:2014fna}
    \bibitem{Khan:2014fna} 
	S.~Khan, N.~A.~Khan and M.~K.~Khan,
  %``Non-maximal Tripartite Entanglement Degradation of Dirac and Scalar fields in Non-inertial frames,''
	Commun.\ Theor.\ Phys.\  {\bf 61}, 281 (2014)
	doi:10.1088/0253-6102/61/3/02
	[arXiv:1402.7152 [quant-ph]].
  %%CITATION = doi:10.1088/0253-6102/61/3/02;%%
  %3 citations counted in INSPIRE as of 30 Oct 2017

%\cite{Lorek}
    \bibitem{Lorek} 
  %``Extraction of genuine tripartite entanglement from the vacuum,''
	K.~Lorek, D.~Pecak, E.~G.~Brown, A.~Dragan,
	Phys. Rev. A 90, 032316 (2014);
	doi:10.1103/PhysRevA.90.032316;
	arXiv:1405.4449.

%\cite{Rota:2015wge}
    \bibitem{Rota:2015wge} 
	M.~Rota,
  %``Tripartite information of highly entangled states,''
	JHEP {\bf 1604}, 075 (2016)
	doi:10.1007/JHEP04(2016)075
	[arXiv:1512.03751 [hep-th]].
  %%CITATION = doi:10.1007/JHEP04(2016)075;%%
  %3 citations counted in INSPIRE as of 28 Oct 2017

%\cite{Bayat:2017bnj}
    \bibitem{Bayat:2017bnj} 
	A.~Bayat,
  %``Scaling of Tripartite Entanglement at Impurity Quantum Phase Transitions,''
	Phys.\ Rev.\ Lett.\  {\bf 118}, no. 3, 036102 (2017)
	doi:10.1103/PhysRevLett.118.036102
	[arXiv:1609.04421 [quant-ph]].
  %%CITATION = doi:10.1103/PhysRevLett.118.036102;%%

%\cite{Krenn2016}
    \bibitem{Krenn2016} 
	M. Krenn, et al.,
  %``Quantum entanglement,''
	Phys.\ Rev.\ Lett.\ {\bf 116}, 865 (2016).

%\cite{Melnikov2018}
    \bibitem{Melnikov2018} 
	A. A. Melnikov, et al.,
	PNAS 115(6), 1221-1226 (2018);
	doi:10.1073/pnas.1714936115;
	arXiv:1706.00868 [quant-ph].

%\cite{Huber2013}
    \bibitem{Huber2013} 
	M. Huber and J. I. de Vicente,
	Phys. Rev. Lett. 110, 030501 (2013);
	doi:10.1103/PhysRevLett.110.030501;
	arXiv:1210.6876 [quant-ph].

%\cite{Malik2016}
    \bibitem{Malik2016} 
	M. Malik, et al.,
  %Manuel Erhard, Marcus Huber, Mario Krenn, Robert Fickler, Anton Zeilinger 
  %Multi-photon entanglement in high dimensions
	Nature Photonics 10, 248-252 (2016);
	doi:10.1038/nphoton.2016.12;
	arXiv:1509.02561.

%\cite{Erhard2018}
    \bibitem{Erhard2018} 
	M. Erhard, R. Fickler, M. Krenn and A. Zeilinger,
	Light: Science \& Applications 7, 17146 (2018);
	doi:10.1038/lsa.2017.146;
	arXiv:1708.06101.

%\cite{Babazadeh2017}
    \bibitem{Babazadeh2017} 
% High-Dimensional Single-Photon Quantum Gates: Concepts and Experiments,
	A. Babazadeh, et al.,
	Phys. Rev. Lett. 119, 180510 (2017);
	doi:10.1103/PhysRevLett.119.180510;
	arXiv:1702.07299.

%\cite{Buniy:2022uwh}
    \bibitem{Buniy:2022uwh}
	R.~V.~Buniy, R.~P.~Feger and T.~W.~Kephart,
  %``Decoherence and the Classes of Maximally Entangled States,''
	doi:10.1142/S0219749924500357
	[arXiv:2210.07618 [quant-ph]].
  %0 citations counted in INSPIRE as of 29 Nov 2022

%\cite{Kauffman}
    \bibitem{Kauffman}
	L.~H.~Kauffman and S.~J.~Lomonaco, arXiv:quant-ph/0205137, quant-ph/0304091, and quant-ph/0403228.

\end{thebibliography}
\end{document}